\documentclass[11pt]{article} 
\usepackage{amsfonts,amssymb,amsmath,amsthm,dsfont,graphicx,hyperref,mathrsfs,euscript,enumitem,bbm}
\usepackage[title, titletoc]{appendix}
\usepackage[onehalfspacing]{setspace}
\usepackage{ifthen,caption,subcaption,lscape}
\hypersetup{pdfborder = {0 0 0},colorlinks=true,linkcolor=blue,urlcolor=blue,citecolor=blue}
\usepackage{natbib}

\usepackage[top=1in,bottom=1in,left=1in,right=1in]{geometry}

	\DeclareMathOperator*{\argmin}{arg\,min}

\renewcommand{\P}{\mathbb{P}}
\newcommand{\E}{\mathbb{E}}

\newcommand{\I}{\mathbbm{1}}

\renewcommand{\c}{c}
\newcommand{\W}{\mathcal{W}}

\begin{document}

% % % % % % % % % % % % % % Title Page % % % % % % % % % % % % % %
\title{\vspace{-0.0in}Regression Discontinuity Designs\thanks{A preliminary version of this manuscript was titled ``Regression Discontinuity Designs: A Review". We are indebted to our collaborators Sebastian Calonico, Max Farrell, Yingjie Feng, Brigham Frandsen, Nicolas Idrobo, Michael Jansson, Luke Keele, Xinwei Ma, Kenichi Nagasawa, Jasjeet Sekhon, and Gonzalo Vazquez-Bare for their continued support and intellectual contribution to our research program on RD designs. We also thank Jason Lindo, Filippo Palomba, Zhuan Pei, and a reviewer for comments. Cattaneo and Titiunik gratefully acknowledge financial support from the National Science Foundation through grants SES-1357561 and SES-2019432, and Cattaneo gratefully acknowledges financial support from the National Institute of Health (R01 GM072611-16). Parts of this article were presented and \href{https://www.youtube.com/watch?v=sUNYMC9ernY&t=56m37s}{video recorded} at the Summer Institute 2021 Methods Lectures of the National Bureau of Economic Research (NBER). General purpose software, replication files, additional articles on RD methodology, and other resources are available at \url{https://rdpackages.github.io/}.}
\bigskip}

\author{Matias D. Cattaneo\thanks{Department of Operations Research and Financial Engineering, Princeton University.} \and
	    Roc\'{i}o Titiunik\thanks{Department of Politics, Princeton University.}}
\maketitle

\begin{abstract}
    The Regression Discontinuity (RD) design is one of the most widely used non-experimental methods for causal inference and program evaluation. Over the last two decades, statistical and econometric methods for RD analysis have expanded and matured, and there is now a large number of methodological results for RD identification, estimation, inference, and validation. We offer a curated review of this methodological literature organized around the two most popular frameworks for the analysis and interpretation of RD designs: the continuity framework and the local randomization framework. For each framework, we discuss three main topics: (i) designs and parameters, which focuses on different types of RD settings and treatment effects of interest; (ii) estimation and inference, which presents the most popular methods based on local polynomial regression and analysis of experiments, as well as refinements, extensions, and alternatives; and (iii) validation and falsification, which summarizes an array of mostly empirical approaches to support the validity of RD designs in practice. 
\end{abstract}


\thispagestyle{empty}

\newpage

\addtocontents{toc}{\protect\thispagestyle{empty}}
\tableofcontents

\thispagestyle{empty}

\newpage

\pagestyle{plain}

\clearpage

\doublespacing
\setcounter{page}{1}
\pagestyle{plain}

% % % % % % % % % % % % % % Beginning of the Document % % % % % % % % % % % % % %
\pagestyle{plain}

% % % % % % % % % % % % % % 
% % % % % % % % % % % % % % 
\section{Introduction}\label{sec: Introduction}

The regression discontinuity (RD) design was introduced by \citet*{Thistlethwaite-Campbell_1960_JEP} as a ``method of testing causal hypotheses'' (p. 317) in settings where the random assignment of treatment is unavailable. The design is applicable in situations where units receive a score, and a treatment is assigned on the basis of a specific rule that links the score with a known cutoff: all units whose score is above the cutoff are assigned to the treatment condition, while all units whose score is below the cutoff are assigned to control condition. Under appropriate assumptions, the discontinuous change in the probability of receiving treatment at the cutoff can be used to learn about causal effects of the treatment on an outcome of interest.

While overlooked at first, the RD design has become one of the most credible non-experimental methods for causal inference and program evaluation. In economics, its use is widespread, driven by the wide applicability of the design across multiple subfields. For example, in development economics, cash transfers are often available to households that reach a threshold on a poverty measure. In labor economics, unemployment insurance or reemployment training is offered to individuals who are older than a certain age or who previously worked a minimum number of hours. In political economy, the incumbent party in two-party majoritarian elections is the candidate that reaches fifty percent of the vote. In public economics, municipalities with populations above certain thresholds are often subjected to different election or tax-sharing rules. RD designs can be employed to study the effect of such policies on various outcomes of interest, including health status, labor supply, fiscal indicators, and saving decisions, among many others. Given that many interventions are not amenable to random experimentation, the RD design offers the opportunity to study treatment effects with a rigor and credibility that would otherwise be unavailable.

We offer a curated review of the methodological literature on the analysis and interpretation of RD designs, providing an update over prior review pieces in economics \citep*{vanderklaauw_2008_Labour,Cook_2008_JoE,Imbens-Lemieux_2008_JoE,Lee-Lemieux_2010_JEL}. Our discussion aims to be comprehensive in terms of conceptual and methodological aspects but omits technical details, which are available in the specific references we give for each topic. We also discuss general implementation issues, but refer the reader to \citet*{Cattaneo-Idrobo-Titiunik_2020_CUP,Cattaneo-Idrobo-Titiunik_2022_CUP} for comprehensive practical guides to empirical analysis. Although we mention examples in various sections, space constraints prevent us from presenting an exhaustive list of empirical applications of RD designs in economics.

We organize our review in three main parts. Section \ref{sec: Designs and Parameters of Interest} introduces the types of RD designs most commonly encountered in practice, and briefly discusses other RD-like research designs. This section focuses on identifiability \citep*{Matzkin_2013_ARE} of the main treatment effects of interest in each setting, and therefore relies on basic restrictions on the underlying data generating process---the statistical model assumed to generate the data. Using the taxonomy proposed by \citet*{Cattaneo-Titiunik-VazquezBare_2017_JPAM}, we consider two distinct conceptual frameworks for RD designs: the \textit{continuity} framework, which relies on limiting arguments via local regression ideas, and the \textit{local randomization} framework, which relies on ideas from the analysis of experiments. Using both frameworks, we provide an overview of canonical RD settings (sharp and fuzzy), multi-dimensional RD designs (multi-cutoff, multi-score, geographic, multiple treatments, and time-varying designs), and other related designs such as kink, bunching, before-and-after, and threshold regression designs. We also discuss issues of internal vs. external validity, and overview methods for extrapolation of RD treatment effects.

In the second part, Section \ref{sec: Estimation and Inference}, we focus on estimation and inference methods for RD designs, presenting the two most common approaches: \textit{local polynomial regression} methods \citep*{Fan-Gijbels_1996_Book}, which are naturally motivated by the continuity framework, and \textit{analysis of experiments} methods \citep*{Rosenbaum_2010_Book,Imbens-Rubin_2015_Book,Abadie-Cattaneo_2018_ARE}, which are naturally motivated by the local randomization framework. For completeness, we also summarize refinements, extensions, and other methods that have been proposed over the years. Building on the two canonical estimation and inference methods for RD designs, we overview the applicability of those methods when the score has discrete support, with either few or many distinct values. To close Section \ref{sec: Estimation and Inference}, we discuss power calculations for experimental design in RD settings, a topic that has lately received renewed attention among practitioners.

In the third and final part, Section \ref{sec: Validation and Falsification}, we focus on methods for validation and falsification of RD designs. These empirical methods are used to provide indirect evidence in favor of the underlying assumptions commonly invoked for RD identification, estimation, and inference. Our discussion includes generic program evaluation methods such as tests based on pre-intervention covariates and placebo outcomes, as well as RD-specific methods such as binomial and density continuity tests, placebo cutoff analysis, donut hole strategies, and bandwidth sensitivity analysis.

We conclude in Section \ref{sec: Conclusion}. General purpose software in \texttt{Python}, \texttt{R}, and \texttt{Stata} for analysis of RD designs, replication files for RD empirical illustrations, further readings, and other resources are available at \url{https://rdpackages.github.io/}.

%%%%%%%%%%%%%%%%%%%%%%%%%%%%%%%%%%%%%%%%%%%%%%%%%%%%%%%
%% SECTION: Designs and Parameters
%%%%%%%%%%%%%%%%%%%%%%%%%%%%%%%%%%%%%%%%%%%%%%%%%%%%%%%
\section{Designs and Parameters}\label{sec: Designs and Parameters of Interest}

The RD design is defined by three key components---a score, a cutoff, and a discontinuous treatment assignment rule---related to each other as follows: all the units in the study are assigned a value of the score (also called a running variable or index), and the treatment is assigned only to units whose score value exceeds a known cutoff (also called threshold). The distinctive feature of the RD design is that the probability of treatment assignment changes from zero to one at the cutoff, and this abrupt change in the probability of being assigned to treatment induces a (possibly smaller) abrupt change in the probability of receiving treatment at the cutoff. Under formal assumptions that guarantee the ``comparability'' of treatment and control units at or near the cutoff, the discontinuous change in the probability of receiving treatment can be used to learn about causal effects of the treatment on outcomes of interest for units with scores at or near the cutoff, because units with scores barely below the cutoff can be used as a comparison (or control) group for units with scores barely above it. The most important threat to any RD design is the possibility that units are able to strategically and precisely change their score to be assigned to their preferred treatment condition \citep*{Lee_2008_JoE,McCrary_2008_JoE}, which might induce a discontinuous change in their observable and/or unobservable characteristics at or near the cutoff and thus confound causal conclusions.

\citet*{Ludwig-Miller_2007_QJE} offer a prototypical empirical application of early RD methodology with a study of the effect of expanding the social services program Head Start on child mortality. In 1965, all U.S. counties whose poverty was above the poverty rate of the 300th poorest county received grant-writing assistance to solicit Head Start funding, while counties below this poverty level did not. The increase in assistance led to an increase in funding for Head Start social services. The authors used an RD design to compare counties barely above and barely below the cutoff to estimate the effect of the increased assistance and funding on child mortality from causes that might be affected by the bundle of Head Start services. In this design, the unit of analysis is the county, the score is the county's poverty rate in 1960, and the cutoff is the poverty rate of the 300th poorest county (which was 59.1984). \citet{Cattaneo-Titiunik-VazquezBare_2017_JPAM} re-analyze this application using the modern methods discussed in Sections \ref{sec: Estimation and Inference} and \ref{sec: Validation and Falsification}. See \cite{Busse-Silva-Risso-Zettelmeyer_2006_AER} for another early application of the RD design in industrial organization.

Unlike other non-experimental methods, RD methodology is only applicable in situations where the treatment is assigned based on whether a score exceeds a cutoff (or where some generalization of the RD rule is employed, as we discuss below). These design-specific features make the RD strategy stand out among observational methods: to study causal RD treatment effects, the score, cutoff and treatment assignment rule must exist ex-ante and be well-defined, with a discontinuous probability of treatment assignment at the cutoff. Importantly, the treatment assignment rule is known and verifiable, and not just postulated or hypothesized by the researcher. These empirically verifiable characteristics give the RD design an objective basis for implementation and validation that is usually lacking in other non-experimental strategies, and thus endow it with superior credibility among non-experimental methods. In particular, the RD treatment assignment rule is the basis for a collection of validation and falsification tests (Section \ref{sec: Validation and Falsification}) that can offer empirical support for the RD assumptions and increase the credibility of RD applications. Furthermore, best practices for estimation and inference in RD designs (Section \ref{sec: Estimation and Inference}) also exploit these design-specific features.

The remaining of this section introduces canonical and extended RD designs, focusing on identifying assumptions for treatment effect parameters of interest employing the potential outcomes framework \citep*{Imbens-Rubin_2015_Book}.

\subsection{Sharp Designs}

RD designs where the treatment assigned and the treatment received coincide for all units are referred to as \textit{sharp}. Sharp RD designs can be used in settings where compliance with treatment is perfect (every unit assigned to treatment receives the treatment and no unit assigned to control receives the treatment), or in cases where compliance with treatment is imperfect (some units assigned to treatment remain untreated and/or viceversa) but the researcher is only interested on the \textit{intention-to-treat} effect of offering treatment. To formalize, suppose that $n$ units, indexed by $i=1,2,\cdots,n$, have a score variable $X_i$. The RD treatment assignment rule is $T_i = \I(X_i\geq c)$ with $T_i$ denoting the observed treatment assignment, $c$ denoting the RD cutoff, and $\I(\cdot)$ denoting the indicator function. Each unit is endowed with two potential outcomes $Y_i(0)$ and $Y_i(1)$, where $Y_i(1)$ is the outcome under the treatment (i.e., when $T_i=1$) and $Y_i(0)$ is the outcome under control (i.e., when $T_i=0$). The probability of treatment assignment changes abruptly at the cutoff from zero to one: $\P[T_i=1|X_i=x] = 0$ if $x<c$ and $\P[T_i=1|X_i=x] = 1$ if $x\geq c$; this is the key distinctive feature of the RD design.

The observed data is $(Y_i,T_i,X_i)'$, $i=1,2,\dots,n$, where $Y_i = T_i Y_i(1)+(1-T_i)Y_i(0)$ captures the fundamental problem of causal inference: $Y_i=Y_i(0)$ for units with $T_i=0$ (i.e. with $X_i<c$) and $Y_i=Y_i(1)$ for units with $T_i=1$ (i.e. with $X_i\geq c$), that is, only one potential outcome is observed for each unit. Crucially, units in the control and treatment groups will not be  comparable in general, as the score $X_i$ often captures important (observable and unobservable) differences between them, and the RD treatment assignment rule induces a fundamental lack of common support in this variable. 

\citet{Hahn-Todd-vanderKlaauw_2001_ECMA} introduced the continuity framework for canonical RD designs. In this framework, potential outcomes are taken to be random variables, with the $n$ units of analysis forming a (random) sample from an underlying population, and the score $X_i$ is assumed to be continuously distributed. Focusing on average treatment effects, the two key identifying assumptions are that (i) the regression functions $\E[Y_i(0)|X_i=x]$ and $\E[Y_i(1)|X_i=x]$ are continuous in $x$ at $c$, and (ii) the density of the score near the cutoff is positive. These assumptions capture the idea that units ``barely below'' and ``barely above'' the cutoff $c$ would exhibit the same average response if their treatment status did not change. Then, by implication, any difference between the average response of treated and control units at the cutoff can be attributed to the treatment and interpreted as the causal average effect of the treatment at the cutoff, that is, for units with score variable $X_i=c$. 

\begin{figure}[t]
		\begin{subfigure}[t]{0.485\textwidth}
			\centering
			\includegraphics[scale=0.42]{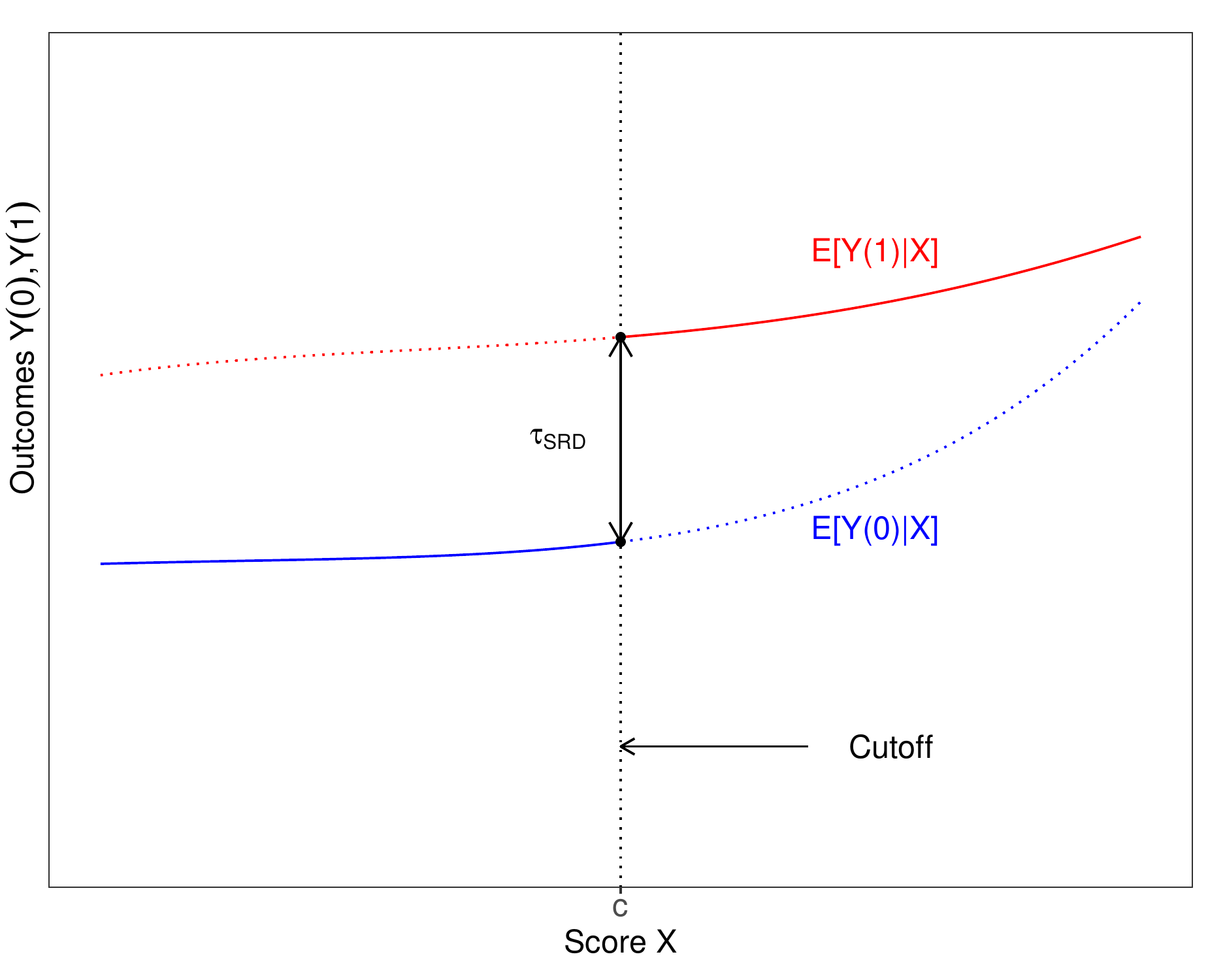}
			\caption{Continuity-based RD Design}\label{subfig:RDcont}	
		\end{subfigure}
		\hspace{0.1in}%
		\begin{subfigure}[t]{0.485\textwidth}
			\centering
			\includegraphics[scale=0.42]{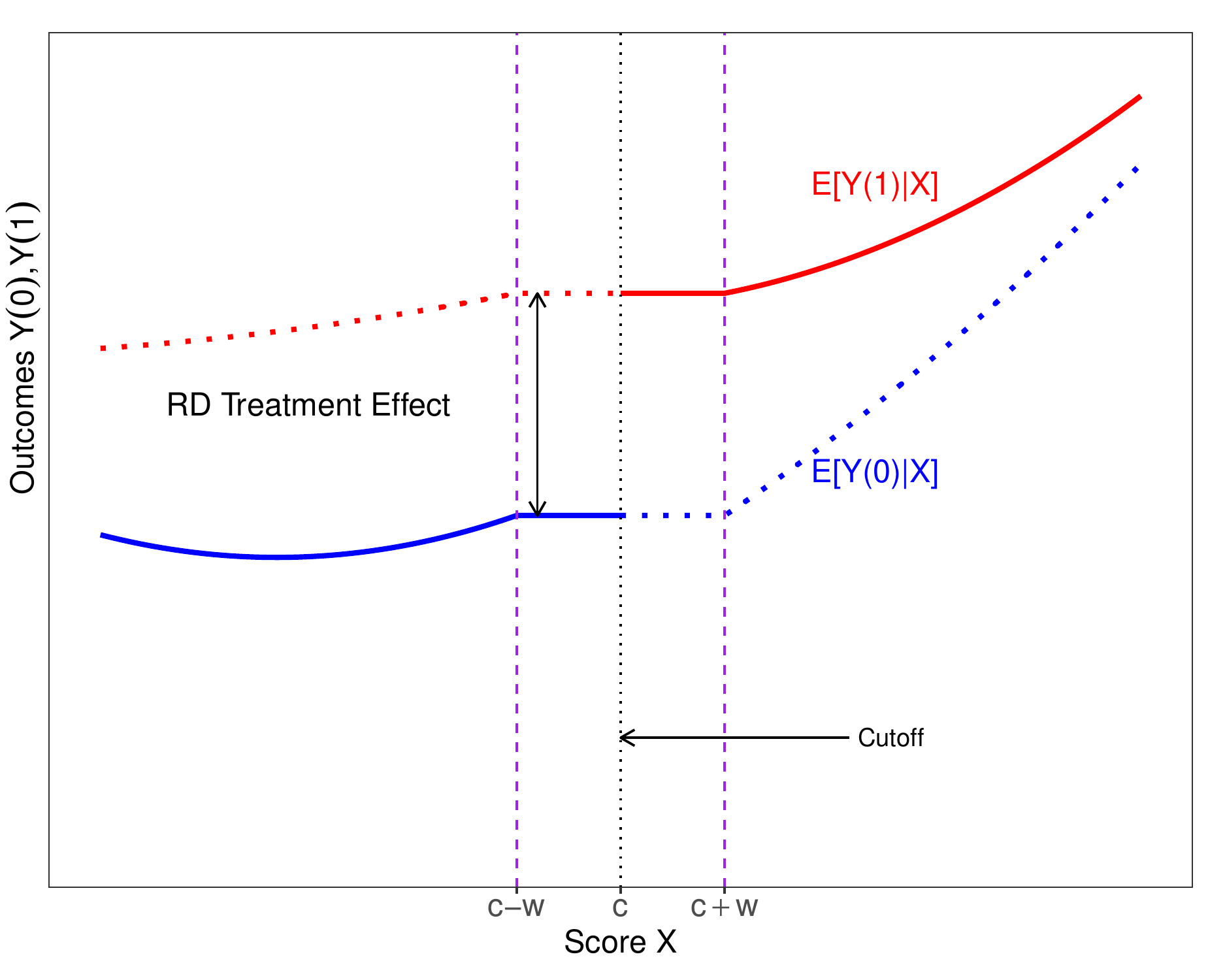}
			\caption{Local Randomization RD Design}\label{subfig:RDlocran}
		\end{subfigure}
		\caption{Schematic Representation of RD designs.}\label{fig:RD}
\end{figure}

Formally, within the continuity-based framework, the canonical sharp RD treatment effect is
\[\tau_\mathtt{SRD} \equiv \E[Y_i(1)-Y_i(0)|X=c] = \lim_{x\downarrow c} \E[Y_i|X_i=x] - \lim_{x\uparrow c} \E[Y_i|X_i=x],\]
which is represented graphically in Figure (\ref{subfig:RDcont}). The identity is the definition of the classical sharp RD parameter, while the equality is the key continuity-based identification result because it links features of the distribution of potential outcomes to features of the distribution of observed outcomes, via a continuity argument. The result asserts that the sharp RD treatment effect is identifiable as the vertical distance at the cutoff between the two conditional expectations, $\lim_{x\uparrow c} \E[Y_i|X_i=x]$ and $\lim_{x\downarrow c} \E[Y_i|X_i=x]$, which can be estimated from data (Section \ref{subsec: Local Polynomial Methods}).

%%% NOTE: Deleted \mu_- and \mu_+ manually from fig-RDCont.pdf -- Will need to change code later to remove that permanently.

The continuity-based identification idea for average sharp RD treatment effects has been extended in several directions. For example, \cite{Frandsen-Frolich-Melly_2012_JoE} study quantile treatment effects, \cite{Xu_2017_JoE} investigates treatment effects using generalized nonlinear models for limited dependent outcome variables, and \cite{Hsu-Shen_2019_JoE} focus on treatment effects conditional on pre-intervention covariates. All these extensions retain the two basic sharp RD features: (i) the treatment assignment is binary and applies to all units in a cross-sectional random sample, and (ii) there is a maximum discontinuity in the probability of treatment at the cutoff, that is, $\lim_{x\downarrow c} \E[T_i|X_i=x] = 1 \neq 0 = \lim_{x\uparrow c} \E[T_i|X_i=x]$. (Recall that $\E[T_i|X_i=x]=\P[T_i=1|X_i=x]$.) In addition, the score variable is assumed to be continuously distributed (Section \ref{subsec: Discrete Score}).

As a complement to the continuity framework, \citet*{Cattaneo-Frandsen-Titiunik_2015_JCI} introduced the local randomization framework for RD analysis. This framework builds on the idea that ``near'' the cutoff the RD design can be interpreted as a randomized experiment or, more precisely, as a natural experiment \citep{Titiunik_2021_HandbookCh}. The heuristic idea of local randomization near the cutoff can be traced back to the original paper of \cite{Thistlethwaite-Campbell_1960_JEP}, who viewed the RD assignment rule as inducing a sort of random assignment near the cutoff and drew a strong analogy between the RD design and a randomized experiment, describing the latter as a design ``for which the regression-discontinuity analysis may be regarded as a substitute'' (p. 310). The analogy between RD designs and randomized experiments took greater force after the work of \citet{Lee_2008_JoE}, who advocated for interpreting the RD design as inducing ``as good as randomized'' (p. 1) variation in treatment status near the cutoff, and also of \citet{Lee-Lemieux_2010_JEL}.

The analogy between RD designs and randomized experiments was first formalized without continuity conditions by \citet{Cattaneo-Frandsen-Titiunik_2015_JCI}. Rather than relying on limits as the score tends to the cutoff and heuristic analogies between units barely above and barely below the cutoff, this framework considers assumptions under which the RD design would produce conditions equivalent to the conditions that would have occurred if a randomized experiment had been conducted in a neighborhood around the cutoff.

The key idea behind the local randomization approach to RD analysis is to assume that units near the cutoff are as-if randomly assigned to treatment. The assumption of as-if random assignment requires the existence of a window $\W=[\c-w,\c+w]$ around the cutoff $c$, assumed symmetric only for simplicity, such that two conditions hold for all units with $X_i\in\W$: (i) the joint probability distribution of scores in $\W$ is known, and (ii) the potential outcomes are not affected by the score in $\W$. Condition (i) assumes that the assignment mechanism of the score is known inside the window. For example, this condition holds when all units have the same probability of receiving all score values in $\W$. Condition (ii) is an exclusion restriction that prevents the potential outcomes (or their distributions if assumed to be random variables) from being a function of the score inside $\W$.
 
The first condition is analogous to the requirement of known assignment mechanism in classical randomized experiments, while the second condition is not typically stated in definitions of randomized experiments because it holds (implicitly) by construction. For example, when a treatment is randomly assigned, this usually involves generating a pseudo-random number and then assigning the treatment or control conditions based on this number. If we interpret this pseudo-random number as a score, it will be unrelated to the potential outcomes by virtue of having been generated independently of any information about the characteristics of the units. In contrast, in an RD design, the score is often correlated with the potential outcomes, and there are multiple channels by which a change in the score, even if small, can induce a change in the potential outcomes. Such a relationship between the score and the potential outcomes would hinder the comparability of units above and below the cutoff within $\W$, because of the lack of common support in the score. For this reason, in a local randomization RD approach, the exclusion restriction inside $\W$ must be made explicit. See \citet{Cattaneo-Titiunik-VazquezBare_2017_JPAM} for a relaxation of this restriction, and \citet{Sekhon-Titiunik_2016_ObsStud,Sekhon-Titiunik_2017_AIE} for more discussion on the differences between randomized experiments and RD designs. \citet*{delacuesta-Imai_2016_ARPS} and \citet*{Cattaneo-Titiunik-VazquezBare_2020_BookCh} further contrast the continuity and local randomization frameworks; see also the related papers in \citet{Cattaneo-Escanciano_2017_AIE}.

The particular formalization of the local randomization framework can take different forms, depending on what approach from the analysis of experiments literature is adopted. The Fisherian approach assumes that the potential outcomes are non-random and the population is finite. The Neyman approach assumes that the potential outcomes are non-random, but sampled from an underlying infinite population. Finally, the so-called super-population approach assumes that the potential outcomes are a random sample. The original formulation of \cite{Cattaneo-Frandsen-Titiunik_2015_JCI} adopted a Fisherian approach, but \citet{Cattaneo-Titiunik-VazquezBare_2017_JPAM} expanded it to include the Neyman and super-population approaches.

The local randomization RD design is illustrated in Figure (\ref{subfig:RDlocran}). The exclusion restriction assumption results in the regression functions $\E[Y_i(1)|X_i=x]$ and $\E[Y_i(0)|X_i=x]$ being constant for all values of the score inside the local randomization neighborhood $\W$; the average treatment effect is therefore the constant vertical distance between these functions inside $\W$. For the $N_\mathcal{W}$ units with $X_i\in\mathcal{W}$, the local randomization sharp RD treatment effect can be therefore defined as
\begin{align*}
    \tau_\mathtt{SLR} &\equiv \frac{1}{N_\mathcal{W}} \sum_{i:X_i\in\mathcal{W}}\E[Y_i(1) - Y_i(0) | X_i\in\mathcal{W}]\\
                      &= \frac{1}{N_\mathcal{W}} \sum_{i:X_i\in\mathcal{W}} \E\Big[ \frac{T_i Y_i}{\P[T_i=1]} \Big| X_i\in\mathcal{W}\Big] 
                       - \frac{1}{N_\mathcal{W}} \sum_{i:X_i\in\mathcal{W}} \E\Big[ \frac{(1-T_i) Y_i}{1-\P[T_i=1]} \Big| X_i\in\mathcal{W}\Big].
\end{align*}
This notation allows for both random and non-random potential outcomes, as well as for countable or uncountable underlying populations, accommodating Fisherian, Neyman, and super-population settings. The estimand $\tau_\mathtt{SLR}$ is identifiable, and can be estimated from observable data as discussed in Section \ref{subsec: Local Randomization Methods}.

Researchers sometimes incorporate additional (pre-intervention) covariates to ``restore'' or ``improve'' the nonparametric identification of sharp RD parameters such as $\tau_\mathtt{SRD}$ or $\tau_\mathtt{SLR}$, depending on the conceptual framework considered. However, as in randomized experiments, it is not possible to restore nonparametric identification of canonical parameters of interest using covariate-adjustment without additional strong assumptions. On the other hand, covariate-adjustment can be used to improve precision in estimation and inference (Section \ref{sec: Estimation and Inference}). Similar considerations apply to the other RD designs discussed below. See \citet*{Calonico-Cattaneo-Farrell-Titiunik_2019_RESTAT} for a detailed discussion, including how to introduce covariates without changing the RD parameters, and \citet{Cattaneo-Keele-Titiunik_2022_HandbookCh} for an overview on covariate-adjusted RD methods.

\subsection{Fuzzy Designs}

Fuzzy RD designs arise in settings where the treatment assigned and the treatment received do not coincide for at least some units, that is, settings where some units assigned to treatment fail to receive the treatment and/or some units assigned to the control condition receive the treatment anyway. To formalize this imperfect compliance setup, we expand the statistical model for sharp RD designs and introduce an additional variable, $D_i$, representing the treatment received. Given the binary assignment $T_i=\I(X_i\geq c)$, each unit now has two binary potential treatments, $D_i(0)$ and $D_i(1)$, where $D_i(1)$ is the treatment received when the unit's score is above the cutoff, and $D_i(0)$ is the treatment received when the unit's score is below the cutoff; the observed treatment is therefore $D_i = D_i(0) (1-T_i) + D_i(1) T_i$. The potential outcomes are generalized to $Y_i(T_i, D_i(T_i))$, leading to four different values depending on the values of $T_i$ and the potential treatments $D_i(T_i)$. For example, $Y_i(0, 1)$ is the outcome that unit $i$ would exhibit when their score is below the cutoff ($T_i=0$) and they receive the treatment anyway ($D_i(0)=1$).

Perfect compliance with treatment assignment means $\P[D_i(0)=0|X_i=x]=1$ for $x<c$ (no units with score below the cutoff receive the treatment) and $\P[D_i(1)=1|X_i=x]=1$ for $x\geq c$ (all units with score below the cutoff receive the treatment), in which case the treatment assignment rule reduces to the sharp RD rule, $D_i = T_i=\I(X_i\geq c)$, and the four potential outcomes reduce to $Y_i(1, 1):=Y_i(1)$ and $Y_i(0, 0):=Y_i(0)$. See \citet{Imbens-Rubin_2015_Book} for a review of this potential outcomes framework.

As mentioned above, if the goal is to learn about the effect of assigning the treatment rather than receiving it, researchers can simply follow a sharp RD analysis where $X_i$ is the score, $T_i$ is seen as the treatment of interest, and $Y_i$ and $D_i$ are the outcomes, which requires invoking standard sharp RD continuity assumptions for the regression functions $\E[Y_i(1,D_i(1))|X_i=x]$, $\E[Y_i(0,D_i(0))|X_i=x]$, $\E[D_i(1)|X_i=x]$, and $\E[D_i(0)|X_i=x]$. 

However, in many cases researchers are also (or instead) interested in the effect of receiving the treatment on the outcome of interest. In general, identification of such treatment effects in settings with imperfect compliance cannot be achieved with the same continuity or local randomization assumptions as in the sharp RD case, and requires additional identifying assumptions. Therefore, learning about interpretable causal treatment effects in fuzzy RD designs is more difficult than in sharp RD designs.

Within the continuity-based framework, the generic fuzzy RD estimand is
\begin{equation*}\label{eq:FRD-est}
\tau_\mathtt{FRD} = \frac{\lim_{x\downarrow c} \E[Y_i|X_i=x] - \lim_{x\uparrow c} \E[Y_i|X_i=x]}{\lim_{x\downarrow c} \E[D_i|X_i=x] - \lim_{x\uparrow c} \E[D_i|X_i=x]}
\end{equation*}
under minimal continuity conditions. The causal interpretation of $\tau_\mathtt{FRD}$ depends on the additional assumptions invoked, and the particular way in which treatment compliance is defined. For example, \citet*{Hahn-Todd-vanderKlaauw_2001_ECMA} showed that if the potential outcomes and the treatment received are independent conditional on the score being near the cutoff, then $\tau_\mathtt{FRD} = \E[Y_i(1)-Y_i(0)|X=c]$, removing the problem of imperfect compliance and identifying the same parameter as in a sharp RD design. However, this local conditional independence assumption is strong and thus not commonly used to interpret the fuzzy RD estimand in a continuity-based approach.

Several authors have derived alternative identification results to interpret $\tau_\mathtt{FRD}$ based on continuity conditions. Many of these approaches seek to characterize the parameter $\tau_\mathtt{FRD}$ as a causal estimand for some subpopulation of ``compliers'' at the cutoff, heuristically defined as units who receive the treatment when they are assigned to the treatment condition and remain untreated when they are assigned to the control condition. The result typically requires a type of monotonicity condition, akin to that invoked in instrumental variables (IV) settings, which rules out units whose compliance decisions are the opposite of their assignment, and is formalized differently depending on the setup. For instance, \cite{Dong_2018_OBES} defines compliance types of units for a given arbitrary value of their score and imposes a monotonicity condition in a neighborhood of the cutoff to achieve identification, while \cite{Arai-Hsu-Kitagawa-Mourifie-Wan_2021_QE} define units' compliance types in a neighborhood near the cutoff, assume that each unit's type is constant for all values of $x$ in this neighborhood, and impose a monotonicity condition at the cutoff. A different definition of types of units is used by \cite{Bertanha-Imbens_2020_JBES}, who assume that, given the value of the cutoff $c$, the decision to take the treatment is only a function of the score via the indicator $\I(X_i \geq c)$, implying that a unit's compliance type is constant for all values of the score with equal treatment assignment. Other related approaches are discussed in \citet{Imbens-Lemieux_2008_JoE}, \cite{Frandsen-Frolich-Melly_2012_JoE}, \citet{Cattaneo-Keele-Titiunik-VazquezBare_2016_JOP}, and \citet{Choi-Lee_2018_EL}.

In addition to monotonicity, the interpretation of $\tau_\mathtt{FRD}$ as the average effect of the treatment requires that the only effect of crossing the cutoff on the outcome occur via the treatment received---a condition analogous to the exclusion restriction in standard IV designs. In other words, conditional on the treatment received, the assignment of the treatment must have no effect on the outcomes. This assumption is violated, for example, when the mere information of having been assigned to the treatment causes participants to change their behavior. In the continuity framework, the exclusion restriction is implied by the continuity assumptions imposed on the underlying conditional expectations at the cutoff.

Regardless of the particular assumptions adopted to achieve a causal interpretation for $\tau_\mathtt{FRD}$, most of these methods arrive at the same qualitative identification result: under continuity of the appropriate functions, and some form of monotonicity, the fuzzy RD estimand $\tau_\mathtt{FRD}$ can be interpreted as the average treatment effect at the cutoff for compliers, where the specific definition of complier varies by approach.

In the local randomization framework, causal interpretability of fuzzy RD treatment effects is better understood because it is analogous to IV results in the analysis of experiments with non-compliance. The generic fuzzy RD estimand can be written as
\[\tau_\mathtt{FLR} = \frac{\vartheta_Y}{\vartheta_D},\]
\[\vartheta_Y = \frac{1}{N_\mathcal{W}} \sum_{i:X_i\in\mathcal{W}} \E\Big[ \frac{T_i Y_i}{\P[T_i=1]} \Big| X_i\in\mathcal{W}\Big] 
              - \frac{1}{N_\mathcal{W}} \sum_{i:X_i\in\mathcal{W}} \E\Big[ \frac{(1-T_i) Y_i}{1-\P[T_i=1]} \Big| X_i\in\mathcal{W}\Big]
\]
\[\vartheta_D = \frac{1}{N_\mathcal{W}} \sum_{i:X_i\in\mathcal{W}} \E\Big[ \frac{T_i D_i}{\P[T_i=1]} \Big| X_i\in\mathcal{W}\Big] 
              - \frac{1}{N_\mathcal{W}} \sum_{i:X_i\in\mathcal{W}} \E\Big[ \frac{(1-T_i) D_i}{1-\P[T_i=1]} \Big| X_i\in\mathcal{W}\Big].
\]
Regardless of the specific local randomization framework, standard ideas from the literature on analysis of experiments can be used to attach a causal interpretation to the estimand $\tau_\mathtt{FLR}$, the most common of which is as an average treatment effect for compliers with scores within the local randomization neighborhood $\W$.

Analogously to the continuity case, if researchers are interested in the effect of the treatment assignment, they can conduct a local randomization sharp RD analysis where $T_i$ is the treatment of interest and $Y_i$ and $D_i$ are the outcomes, which requires making local randomization assumptions for the potential outcomes as well as the potential treatments.

\subsection{Multi-Dimensional Designs}

In the canonical RD designs discussed so far, both the score and the cutoff are one-dimensional, and the treatment is binary. We now consider generalizations of these conditions, which we generically label \textit{multi-dimensional} RD designs. These generalized RD designs require specific modifications for interpretation and analysis. Conceptually, however, the core of the design remains the same: a discontinuous change in the probability of receiving treatment is the basis for comparing units assigned to treatment and control near the region where such a change occurs, under assumptions that guarantee the comparability of those units. Most of the designs discussed in this subsection admit sharp and fuzzy versions, although we avoid distinguishing each case due to space limitations.

\citet*{Papay-Willett-Murnane_2011_JoE} introduced the Multi-Score RD design; see also \citet*{Reardon-Robinson_2012_JREE} and \citet*{Wong-Steiner-Cook_2013_JEBS}. In its more general version, the Multi-Score RD design includes two or more scores assigning units to a range of different treatment conditions, but a more common setup is one where the score is multi-dimensional but the treatment is still binary. For example, students may need to take a language exam and a mathematics exam, and obtain a high enough grade in both exams to be admitted to a school. In such two-dimensional example, the score is the vector $\mathbf{X}_i = (X_{1i}, X_{2i})'$, the cutoff is the vector $\mathbf{c}=(c_1,c_2)$, and the treatment assignment rule is $D_i = \I(X_{1i} \geq c_1)\I(X_{2i} \geq c_2)$. This rule shows that there are potentially infinite ways for a unit to be barely treated or barely control, that is, there are infinitely many points at which the treatment assignment jumps discontinuously from zero to one, leading to the definition of a treatment effect curve. This idea generalizes to higher-dimensions.

An important particular case of the Multi-Score RD design is the Geographic RD design, where the treatment is assigned to units that reside in a particular geographic area (the \textit{treated area}), and not assigned to units who are in an adjacent area (the \textit{control area}). The RD score is two-dimensional to reflect each unit's position in space, usually latitude and longitude. The geographic boundary that separates the treated and control areas thus forms an infinite collection of points at which the treatment assignment jumps discontinuously from zero to one. See \citet*{Black_1999_QJE} for a pioneer use of a Geographic RD research design, and \cite{Dell_2010_ECMA} for the first paper to explicitly exploit latitude and longitude to achieve RD-specific geographic identification of an average treatment effect. Modern analysis of Geographic RD designs is discussed by \cite{Keele-Titiunik_2015_PA} using continuity-based ideas, and by \citet*{Keele-Titiunik-Zubizarreta_2015_JRSSA} using local randomization ideas. \citet*{Keele-etal_2017_AIE} discuss available options for analysis when the exact geographic location of each unit in the dataset is unavailable.

\cite{Cattaneo-Keele-Titiunik-VazquezBare_2016_JOP} introduced the Multi-Cutoff RD design, where different units in the study receive the treatment according to different cutoff values along a univariate score. In this setting, the variable $X_i$ continues to denote the scalar score for unit $i$, but instead of a fixed common cutoff the setting now allows for an additional random variable, $C_i$, that denotes the cutoff faced by each unit. The cutoff random variable $C_i$ has support $\mathcal{C}$, which can be countable or uncountable. For example, a natural assumption is $\mathcal{C}=\{c_1,c_2,...,c_J\}$, where the values $c_1,c_2,...,c_J$ can either be distinct cutoffs in an RD design or a discretization of a continuous support. If $\mathcal{C}=\{c\}$, then the canonical (sharp and fuzzy) RD designs are recovered. 

The Multi-Score RD design can be recast and analyzed as a Multi-Cutoff RD design by appropriate choice of the score and cutoff variables, $X_i$ and $C_i$. For example, it is common practice to study Multi-Score RD treatment effects for a finite set of values (i.e., a discretization) of the treatment assignment curve, which can be naturally interpreted as a Multi-Cutoff RD design with each cutoff being one point on the discretization of the curve. In addition, as discussed in Section \ref{subsec: Extrapolation}, Multi-Cutoff RD designs may be useful for extrapolation of RD treatment effects \citep*{Cattaneo-Keele-Titiunik-VazquezBare_2021_JASA}.

Both Multi-Score and Multi-Cutoff RD designs offer a rich class of RD treatment effects. For a given treatment assignment boundary curve (in the Multi-Score RD design) or each cutoff value (in the Multi-Cutoff RD design), a sharp or fuzzy RD treatment effect can be defined, using either the continuity or local randomization frameworks. Furthermore, it is common practice to also normalize and pool the data along the treatment assignment boundary curve or the multiple cutoff values in order to consider a single, pooled RD treatment effect. \cite{Keele-Titiunik_2015_PA} and \citet*{Cattaneo-Keele-Titiunik-VazquezBare_2016_JOP} discuss a causal link between the normalized-and-pooled RD treatment effect and the RD treatment effects along the treatment assignment curve or at multiple cutoffs. See \citet{Cattaneo-Idrobo-Titiunik_2022_CUP} for a practical introduction to Multi-Score and Multi-Cutoff RD designs methods.

In addition to the Multi-Cutoff and Multi-Score RD designs, there are several other types of RD designs with multi-dimensional features. \citet*{Dong-Lee-Gou_2021_JASA} consider an RD design with a continuous (rather than binary or multi-valued) treatment, within the continuity framework, and show that RD causal effects can be identified based on changes in the probability distribution of the continuous treatment at the cutoff. \citet{Caetano-Caetano-Escanciano_2021_wp} investigate continuity-based identification of RD causal effects where the treatment is multi-valued with finite support. \citet*{Grembi-Nannicini-Troiano_2016_AEJ-Pol} introduce a second dimension to the RD design in a setup where there are two time periods, and discuss continuity-based identification of RD treatment effects across the two periods, in a design they call \textit{difference-in-discontinuities} design, building on ideas from the difference-in-differences design in program evaluation. \citet*{Cellini-Ferreira-Rothstein_2010_QJE} consider a different type of multi-dimensionality induced by a dynamic context in which the RD design occurs in multiple periods for the same units and the score is re-drawn in every period so that a unit may be assigned to treatment in one period but control in future periods; see also \citet{Hsu-Shen_2021_wp} for an econometric analysis of a dynamic RD design within the continuity-based framework. \citet*{Lv-Sun-Lu-Li_2019_EL} consider the generalization of the RD design to survival data settings, where the treatment is assigned at most once per unit and the outcome of interest is the units' survival time in a particular state, which may be censored.  \cite{Xu_2018_JoE} also studies an RD designs with duration outcomes, but assumes the outcomes have discrete support.

\subsection{Extrapolation}\label{subsec: Extrapolation}

RD designs offer credible identification of treatment effects at or near the cutoff, via either continuity or local randomization frameworks, often with superior internal validity relative to other observational methods. However, an important limitation of RD designs is that they have low external validity. In the absence of additional assumptions, it is not possible to learn about treatment effects away from the cutoff---that is, to extrapolate the effect of the treatment. In the continuity-based sharp RD design, for example, the average treatment effect at the cutoff is $\tau_\mathtt{SRD}=\tau_\mathtt{SRD}(c)$ with $\tau_\mathtt{SRD}(x) = \E[Y_i(1)-Y_i(0)|X_i=x]$. However, researchers and policy-makers may also be interested in average treatment effects at other values of the score variable such as $\tau_\mathtt{SRD}(x)$ for $x$ different (and possibly far) from $c$, which are not immediately available.

In recent years, several approaches have been proposed to enable valid extrapolation of RD treatment effects. In the context of sharp designs, \citet*{Mealli-Rampichini_2012_JRSSA} and \citet*{Wing-Cook_2013_JPAM} rely on an external pre-intervention measure of the outcome variable and employ parametric methods to impute the treated-control differences of the post-intervention outcome above the cutoff. \citet{Dong-Lewbel_2015_ReStat} and \citet{Cerulli-Dong-Lewbel-Poulsen_2017_AIE} study local extrapolation methods via derivatives of the RD average treatment effect in the continuity-based framework, while \citet{Angrist-Rokkanen_2015_JASA} employ pre-intervention covariates under a conditional ignorability condition within a local randomization framework. \citet*{Rokkanen_2015_wp} relies on multiple measures of the score, which are assumed to capture a common latent factor, and employs ideas from a local randomization framework to extrapolate RD treatment effects away from the cutoff assuming that the potential outcomes are conditionally independent of the available measurements given the latent factor. In the context of continuity-based fuzzy designs, \citet{Bertanha-Imbens_2020_JBES} exploit variation in treatment assignment generated by imperfect compliance imposing independence between potential outcomes and compliance types to extrapolate average RD treatment effects.

In the context of (sharp and fuzzy) Multi-Cutoff RD designs, and from both continuity and local randomization perspectives, \citet*{Cattaneo-Keele-Titiunik-VazquezBare_2016_JOP} discuss extrapolation in settings with ignorable cutoff assignment, while \citet*{Cattaneo-Keele-Titiunik-VazquezBare_2021_JASA} discuss extrapolation allowing for heterogeneity across cutoffs that is invariant as function of the score. In addition, within the continuity framework, \citet*{Bertanha_2020_JoE} discusses extrapolation of average RD treatment effects assuming away cutoff-specific treatment effect heterogeneity. In the specific context of geographic RD designs, \citet*{Keele-Titiunik_2015_PA} and \citet{Keele-Titiunik-Zubizarreta_2015_JRSSA} discuss extrapolation of average treatment effects along the geographic boundary determining treatment assignment. See also \citet*{Galiani-McEwan-Quistorff_2017_AIE} and \citet*{Keele-etal_2017_AIE} for related discussion.

Finally, there is a literature that investigates the relationship between RD designs and randomized experiments and discusses the relationship between local RD effects and more general parameters.  For example, \citet{Cook-Wong_2008_AES} report a study of three empirical RD designs and how they compare to experimental benchmarks created by randomized experiments of the same intervention. See \citet{Chaplin-etal_2018_JPAM}, \citet{Hyytinen-etal_2018_QE}, and \citet{DeMagalhaes-etal_2020_wp} for more recent results on RD recovery of experimental benchmarks.

\subsection{Other RD-like Designs}

The are several other research designs with qualitatively similar features to RD designs but also important differences. Some of these research designs can be analyzed using the econometric tools discussed in Sections \ref{sec: Estimation and Inference} and \ref{sec: Validation and Falsification}, while for others specific modifications would be needed.

\citet*{Card-Lee-Pei-Weber_2015_ECMA,Card-Lee-Pei-Weber_2017_AIE} introduce the regression kink (RK) design. Canonical and generalized RD designs assume a discontinuous jump in the probability of receiving treatment at the cutoff. In contrast, in an RK design there is no such discontinuous jump; instead, the assignment rule that links the policy (or treatment) and the score is assumed to change slope at a known cutoff point (also known as the kink point). The expectation is that the regression function of the observed outcome will be continuous at all values of the score, but its slope will be discontinuous at the cutoff point where the policy rule has the kink. Such kinks arise naturally in social and behavioral sciences, such as when a tax or benefit schedule is a piecewise linear function of baseline income. The core idea in the RK design is to examine the slope of the conditional relationship between the outcome and the score at the kink point in the policy formula. \citet*{Chiang-Sasaki_2019_JoE} extends this idea to quantile treatment effects. See also \citet{Dong_2018_wp} for related discussion. From the point of view of implementation, estimation and inference of kink treatment effects is equivalent to estimating RD treatment effects for derivatives of regression functions. In the RD estimation and inference literature, generic estimands defined as differences of first derivatives of regression functions at the cutoff, or ratios thereof, are referred to as kink RD designs \citep*[e.g.,][Sections 3.1 and 3.3]{Calonico-Cattaneo-Titiunik_2014_ECMA}.

Another area with interesting connections to the RD design is the literature on bunching and density discontinuities \citep{Kleven_2016_ARE,Jales-Yu_2017_AIE}. In this setting, the objects of interest are related to discontinuities and other sharp changes in probability density functions, usually motivated by economic or other behavioral models \citep[e.g.][]{Bajari-Hong-Park-Town_2011_wp}. These models typically exploit discontinuities on treatment assignment or policy rules, like RD or RK designs, but they focus on estimands that are outside the support of the observed data. As a consequence, identification (as well as estimation and inference) requires additional parametric modelling assumptions that are invoked for extrapolation purposes. See \citet*{Blomquist-Newey-Kumar-Liang_2021_JPE} for a discussion of nonparametric identification in bunching and density discontinuities settings. 

Before-and-after (or event) studies are also sometimes portrayed as ``RD designs in time'' where the time index is taken as the score variable and the analysis is conducted using RD methods \citep{Hausman-Rapson_2018_ARRE}. While it is generally hard to reconcile those event studies as traditional RD designs, the local randomization RD framework can sometimes be adapted and used to analyze RD designs in time (see Sections \ref{subsec: Local Randomization Methods} and \ref{subsec: Discrete Score} for more discussion). Another RD-like example is \citet{Porter-Yu_2015_JoE}, who consider settings where the cutoff in an assignment rule is unknown and must be estimated from the data, a problem that can occur, for example, when unknown preferences generate tipping points where behavior changes discontinuously \citep{Card-Mas-Rothstein_2008_QJE}. \citet{Hansen_2017_JBES} explores a similar problem in the RK context.

%%%%%%%%%%%%%%%%%%%%%%%%%%%%%%%%%%%%%%%%%%%%%%%%%%%%%%%
%% SECTION: Estimation and Inference
%%%%%%%%%%%%%%%%%%%%%%%%%%%%%%%%%%%%%%%%%%%%%%%%%%%%%%%
\section{Estimation and Inference}\label{sec: Estimation and Inference}

We now discuss methods for RD analysis, focusing on the continuity-based  and local randomization approaches separately. Before presenting formal methods for estimation and inference, we briefly discuss how to present the RD design visually using a combination of global and local techniques.

\subsection{Visualization of the RD design}

One of the advantages of the RD design is its transparency. The running variable gives the observations a natural ordering, and the treatment assignment rule links this ordering to treatment assignment, which in turn is hypothesized to affect an outcome variable of interest. These ingredients can all be displayed graphically to provide a visual illustration of the design, usually referred to as an RD Plot. This plot has become a central component of RD empirical analysis, regardless of the formal methodology employed for estimation and inference. Although a graphical illustration cannot replace formal econometric analysis, it is often helpful to display the general features of the study, to gauge whether the discontinuity observed at the cutoff is unusual with respect to the overall shape of the regression function, as well as to have an overall understanding of the data to be used in the subsequent formal analysis.

The RD plot graphs different estimates of the conditional expectation of the outcome function given the score, $\E[Y_i|X_i=x]$, as a function of $x$. The typical plot has two ingredients: a global polynomial fit of the outcome on the score, and local sample means of the outcome computed in small bins of the score variable. Both ingredients are calculated separately for treated and control observations because $\E[Y_i|X_i=x]$ may exhibit a discontinuity at the cutoff, thereby capturing two different regression functions: $\E[Y_i(0)|X_i=x]$ for observations below the cutoff, and $\E[Y_i(1)|X_i=x]$ for observations above the cutoff (Figure \ref{fig:RD}).

Each ingredient plays a different role. The two global polynomial fits are designed to show a smooth global approximation of the regression functions that gives a sense of their overall shape---for example, to see whether they are increasing or decreasing in $x$, or whether there are nonlinearities. The typical strategy is to fit a polynomial of order four, but this should be modified as needed. The local means are constructed by partitioning the support of the score into disjoint bins, and computing sample averages of the outcome for each bin, using only observations whose score value falls within that bin; each local mean is plotted against the bin mid-point. In contrast to the global fit, the local means are intended to give a sense of the local behavior of the regression function. In this sense, by simultaneously depicting two representations of the regression function, one smooth (the global polynomial fit) and the other discontinuous (the local means), the RD plot provides rich information about the overall data underlying the RD design.

\citet*{Calonico-Cattaneo-Titiunik_2015_JASA} propose methods for implementing RD plots in a data-driven and disciplined way, and \citet[Section 3]{Cattaneo-Idrobo-Titiunik_2020_CUP} offer a practical introduction to RD Plots. In addition, see \citet*{Pei-Lee-Card-Weber_2021_JBES} for a discussion on how to choose polynomial orders in RD designs, which could be used to further improve the implementation of RD Plots.

Importantly, RD plots should not be used as substitutes for formal estimation and inference of RD effects. In particular, sometimes the global polynomial fit will exhibit a ``jump'' at the cutoff, which might suggest the presence of a treatment effect. Although useful as graphical and suggestive evidence, such jump should not be interpreted as the definitive RD effect, as higher-order polynomial approximations are poorly behaved at or near boundary points---a problem known as the Runge's phenomenon in approximation theory \citep*[Section 2.1.1]{Calonico-Cattaneo-Titiunik_2015_JASA}. Furthermore, global polynomials for RD estimation and inference lead to several other  undesirable features \citep{Gelman-Imbens_2019_JBES}, and are therefore not recommended for analysis beyond visualization. The limitations of visual RD analysis are investigated experimentally by \cite{Korting-etal_2021_visual}, who find that changing the specification of RD plots while keeping the underlying model constant leads participants to draw different conclusions.

%%%%%%%%%%%%%%%%%%%%%%%%%%%%%%%%%%%%%%%%%%%%%%%%%%%%%%%
%% SUBSECTION: Local Polynomial Methods
%%%%%%%%%%%%%%%%%%%%%%%%%%%%%%%%%%%%%%%%%%%%%%%%%%%%%%%
\subsection{Local Polynomial Methods}\label{subsec: Local Polynomial Methods}

The continuity-based identification result of \cite{Hahn-Todd-vanderKlaauw_2001_ECMA} naturally justifies estimation of RD effects via local polynomial nonparametric methods \citep{Fan-Gijbels_1996_Book}. These methods have become the standard for estimation and inference in the RD design under continuity conditions since the early work of \citet{Porter_2003_wp} and \citet{Sun_2005_wp}, and the more recent work of \citet{Imbens-Kalyanaraman_2012_REStud} and \citet{Calonico-Cattaneo-Titiunik_2014_ECMA}. Local polynomial methods are preferable to global polynomial methods because, as mentioned above, they avoid several of the methodological problems created by the use of global polynomials such as erratic behavior near boundary points, counterintuitive weighting, overfitting, and general lack of robustness. 

Local polynomial analysis for RD designs is implemented by fitting $Y_i$ on a low-order polynomial expansion of $X_i$, separately for treated and control observations, and in each case using only observations near the cutoff rather than all available observations, as determined by the choice of a kernel function and a bandwidth parameter. The implementation requires four steps. First, choose the polynomial order $p$ and the weighting scheme or kernel function $K(\cdot)$. Second, given the choices of $p$ and $K(\cdot)$, choose a bandwidth $h$ that determines a neighborhood around the cutoff so that only observations with scores within that neighborhood are included in the estimation. Third, given the choices of $p$, $K(\cdot)$, and $h$, construct point estimators using standard least-squares methods. Fourth, given the above steps, conduct valid statistical inference for the RD parameter of interest. We discuss these steps in more detail in the remaining of this subsection, focusing on methodological aspects.

The order of the polynomial should always be low, to avoid overfitting and erratic behavior near the cutoff point. The default recommendation is $p=1$ (local linear regression), but $p=2$ (local quadratic regression) or $p=3$ (local cubic regression) are also reasonable choices in some empirical settings. See \citet*{Pei-Lee-Card-Weber_2021_JBES} for a data-driven choice of $p\in\{0,1,2,3\}$ based on minimizing the asymptotic mean squared error (MSE) of the RD point estimator. The choice of kernel, $K(\cdot)$, assigns different weights for different observations according to the proximity of each observation's score to the cutoff. Common choices are a triangular kernel which assigns weights that are highest at the cutoff and decrease linearly for values of the score away from the cutoff, and a uniform kernel that assigns the same weight to all observations. The triangular kernel has a point estimation MSE optimality property when coupled with an MSE-optimal bandwidth choice (discussed below), while the uniform kernel has an inference optimality property (i.e., it minimizes the asymptotic variance of the local polynomial estimator, and hence the interval length of commonly used confidence intervals). Both choices of kernel function are reasonable, although in practice the triangular kernel is often the default.

Once $p$ and $K(\cdot)$ have been selected, the researcher must choose the bandwidth. When a particular value of $h$ is chosen, this means that the polynomial fitting will only use observations with $X_i \in [c-h,c+h]$, since the kernel functions used in RD designs are always compact supported. The notation assumes that the same bandwidth $h$ is used both above and below the cutoff for simplicity, but several approaches have been proposed to accommodate distinct bandwidths for control and treatment groups. The choice of bandwidth is critical because empirical results are generally sensitive to changes in this tuning parameter, and therefore choosing the bandwidth in an arbitrary manner is discouraged. Instead, the recommendation is always to use a data-driven procedure that is optimal for a given criterion, leading to a transparent, principled, and objective choice for implementation that ultimately removes the researcher's discretion. Such an approach to tuning parameter selection enhances replicability and reduces the potential for p-hacking in empirical work.

\cite{Imbens-Kalyanaraman_2012_REStud} first proposed to choose $h$ to minimize a first-order approximation to the MSE of the RD point estimator, leading to an MSE-optimal bandwidth choice. For implementation, they use a first-generation plug-in rule, where unknown quantities in the MSE-optimal bandwidth formula are replaced by inconsistent estimators. \citet{Calonico-Cattaneo-Titiunik_2014_ECMA} later obtained more general MSE expansions and MSE-optimal bandwidth choices, which are implemented using a second-generation plug-in rule (i.e., unknown quantities in the MSE-optimal bandwidth formula are replaced by consistent estimators thereof). \citet*{Calonico-Cattaneo-Farrell_2020_ECTJ} take a different approach and develop optimal bandwidth choices for inference: their proposed bandwidth choices either (i) minimize the coverage error (CE) of confidence intervals or (ii) optimally trade off coverage error and length of confidence intervals. The CE-optimal bandwidth choice can also be interpreted as optimal to reduce the higher-order errors in the distributional approximation used for local polynomial statistical inference. \citet{Cattaneo-VazquezBare_2016_ObsStud} provide an overview of bandwidth selection methods for RD designs.

Given a polynomial order $p$, a kernel function $K(\cdot)$, and a bandwidth $h$, the RD local polynomial point estimator is obtained by fitting two weighted least squares regressions of $Y_i$ on a polynomial expansion of $X_i$ as follows:
\begin{align*}
\boldsymbol{\widehat{\beta}}_- = \argmin_{b_0, \ldots, b_p} \sum_{i=1}^n \I(X_i < \c)\left( Y_i - b_{0} - b_{1} (X_i - \c) -  b_{2} (X_i - \c)^2 - \ldots - b_{p} (X_i - \c)^p \right)^2 K\Big(\frac{X_i - c}{h}\Big)
\end{align*}
and
\begin{align*}
\boldsymbol{\widehat{\beta}}_+ = \argmin_{b_0, \ldots, b_p} \sum_{i=1}^n \I(X_i \geq \c)\left( Y_i - b_{0} - b_{1} (X_i - \c) -  b_{2} (X_i - \c)^2 - \ldots - b_{p} (X_i - \c)^p \right)^2 K\Big(\frac{X_i - c}{h}\Big),\end{align*}
where $\boldsymbol{\widehat{\beta}}_-  = (\widehat{\beta}_{-,0}, \widehat{\beta}_{-,1},\ldots, \widehat{\beta}_{-,p})'$ and $\boldsymbol{\widehat{\beta}}_+  = (\widehat{\beta}_{+,0}, \widehat{\beta}_{+,1},\ldots, \widehat{\beta}_{+,p})'$ denote the resulting least squares estimates for control and treatment groups, respectively. The RD point estimator of the sharp treatment effect $\tau_\mathtt{SRD}$ is the estimated vertical distance at the cutoff, that is, the difference in intercepts: $\widehat{\tau}_{\mathtt{SRD}}(h) = \widehat{\beta}_{+,0}-\widehat{\beta}_{-,0}$. Under standard assumptions, $\widehat{\tau}_{\mathtt{SRD}}(h)$ will be consistent for $\tau_{\mathtt{SRD}}= \E[Y_i(1)-Y_i(0)|X=c]$. It is common practice to use $\widehat{\tau}_{\mathtt{SRD}}(h_\mathtt{MSE})$, where $h_\mathtt{MSE}$ denotes an MSE-optimal bandwidth choice, thereby delivering not only a consistent but also an MSE-optimal point estimator.

A crucial question is how to perform valid statistical inference for the RD parameter based on local polynomial methods with an MSE-optimal bandwidth. At first sight, it seems that standard least squares results could be applied to construct confidence intervals because, given a bandwidth choice, estimation is implemented via weighted least-squares linear regression. However, conventional least squares inference methods are parametric, assuming the polynomial expansion and kernel function used capture the correct functional form of the unknwon conditional expectations. Conventional least squares methods thus assume away any misspecification error to conclude that the usual t-statistic has an asymptotic standard normal distribution, which (if true) would lead to the standard approximately $95\%$ least squares confidence interval:
\[I_\mathtt{LS}=\Big[  ~\widehat{\tau}_{\mathtt{SRD}}(h_\mathtt{MSE})~\pm~1.96\cdot\sqrt{\widehat{\mathsf{V}}} ~\Big],\]
where $\sqrt{\widehat{\mathsf{V}}}$ denotes any of the usual standard error estimators in least-squares regression settings.

However, RD local polynomial methods are nonparametric in nature and thus view the polynomial fit near the cutoff as an approximation to unknown regression functions, which generally carries a misspecification error (or smoothing bias). In fact, if there was no misspecification bias, then standard MSE-optimal bandwidth choices would not be well-defined to begin with. It is therefore not possible to simultaneously rely on MSE-optimal bandwidth choices and conduct inference without taking into account misspecification errors in the approximation of the conditional expectations near the cutoff. Crucially, the approximation errors will translate into a bias in the distributional approximation of the RD local polynomial estimator, where in general we have
\[\frac{\widehat{\tau}_{\mathtt{SRD}}(h_\mathtt{MSE}) - \tau_\mathtt{SRD}}{\sqrt{\widehat{\mathsf{V}}}}\approx_{d}\mathcal{N}(\mathsf{B},1),\]
with $\approx_{d}$ denoting approximately in distribution, $\mathcal{N}(\cdot)$ denoting the Gaussian distribution, and $\mathsf{B}$ capturing the (standardized) misspecification bias or approximation error. Ignoring the misspecification bias can lead to substantial over-rejection of the null hypothesis of no treatment effect ($\mathsf{H}_0:\tau_\mathtt{SRD}=0$). For example, back-of-the-envelop calculations show that a nominal $95\%$ confidence interval would have an empirical coverage of about $80\%$.

For some choices of $h$ smaller than $h_\mathtt{MSE}$, the bias term $\mathsf{B}$ will vanish asymptotically, which is called ``undersmoothing'' in the nonparametric literature. Although researchers could choose some $h<h_{\mathtt{MSE}}$ and appeal to asymptotics, such approach is undisciplined and arbitrary (the bias may still be large in finite samples), leads to reduced power (fewer observations are used for inference), and requires employing different observations for MSE-optimal point estimation and for valid inference. As a principled alternative, \citet*{Calonico-Cattaneo-Titiunik_2014_ECMA} propose to use a different confidence interval that estimates and removes the misspecification bias in the asymptotic approximation and, at the same time, adjusts the standard error formula to  account for the additional variability introduced in the bias-estimation step. Specifically, the approximately $95\%$ \textit{robust bias-corrected} confidence interval is:
\[I_{\mathtt{RBC}} = \left[ ~\Big(\widehat{\tau}_{\mathtt{SRD}}(h_\mathtt{MSE}) - \widehat{\mathsf{B}} \Big) ~\pm~ 1.96\cdot\sqrt{\widehat{\mathsf{V}} + \widehat{\mathsf{W}}} ~\right],\]
where $\widehat{\mathsf{B}}$ denotes the estimated bias correction and $\widehat{\mathsf{W}}$ denotes the adjustment in the standard errors. 

The robust bias-corrected confidence interval $I_\mathtt{RBC}$ is recentered and rescaled with respect to the conventional interval $I_\mathtt{LS}$. The new standardization is theoretically justified by a more general large-sample distributional approximation of the bias-corrected estimator that explicitly accounts for the potential contribution of bias correction to the finite-sample variability of the usual t-statistic. This confidence interval is valid whenever the conventional confidence interval is valid, and continues to offer correct coverage in large samples even when the conventional confidence interval does not---hence the name \textit{robust}. In particular, $I_\mathtt{RBC}$ is valid even when the MSE-optimal bandwidth $h_{\mathtt{MSE}}$ is employed, thereby allowing researchers to use the same bandwidth for both (optimal) point estimation and (valid, possibly optimal) inference. See \citet*[Section 4]{Cattaneo-Idrobo-Titiunik_2020_CUP} for a practical guide.

Robust bias correction (RBC) inference methods have been further developed in recent years. From the theoretical side, \citet{Kamat_2018_ET} gives conditions under which RBC inference has uniform validity, \citet{Calonico-Cattaneo-Farrell_2018_JASA,Calonico-Cattaneo-Farrell_2022_Bernoulli} demonstrate superior higher-order properties of RBC inference both pointwise and uniformly over standard data generating classes, and \citet{Tuvaandorj_2020_JoE} establishes minimax optimality. Methodologically, these theoretical results justify a simple implementation of RBC inference where the point estimation polynomial order $p$ is increased when conducting inference without requiring any other changes \citep*[Remark 7]{Calonico-Cattaneo-Titiunik_2014_ECMA}, which allows for using the same data for MSE-optimal point estimation and valid inference. Finally, from a practical perspective, RBC inference has been found to exhibit excellent performance in both simulations and replication studies \citep*[e.g.,][]{Ganong-Jager_2018_JASA,Hyytinen-etal_2018_QE,DeMagalhaes-etal_2020_wp}.

The estimation and inference methods described so far extend naturally to fuzzy RD designs, where the local polynomial point estimator is now a ratio of sharp RD estimators, and to sharp and fuzzy kink RD designs where the point estimator is based on estimated higher-order derivatives at the kink point. For example, the fuzzy RD estimand $\tau_{\mathtt{FRD}}$ is estimated by $\widehat{\tau}_{\mathtt{FRD}}(h) = \widehat{\tau}_{\mathtt{SRD}}(h) / \widehat{\varrho}_{\mathtt{SRD}}(h)$ with $\widehat{\varrho}_{\mathtt{SRD}}(h)$ denoting a sharp RD local polynomial estimate using the observed treatment  $D_i$ as outcome instead of $Y_i$---that is, $\widehat{\varrho}_{\mathtt{SRD}}(h)$ is the sharp RD effect on receiving the treatment at the cutoff. For more details on fuzzy and higher-order derivative estimation and inference see \citet{Calonico-Cattaneo-Titiunik_2014_ECMA}, and \citet*[Section 3]{Cattaneo-Idrobo-Titiunik_2022_CUP} for a practical discussion.

The estimation and inference results for canonical sharp and fuzzy RD designs have been expanded in multiple directions. \citet*{Arai-Ichimura_2016_EL,Arai-Ichimura_2018_QE} propose optimal MSE-optimal bandwidth selection for different control and treatment bandwidths and develop RBC inference, \citet{Bartalotti-Calhoun-He_2017_AIE} and \citet{He-Bartalotti_2020_EJ} investigate resampling methods and their connection with RBC inference, and \citet{Bartalotti-Brummet_2017_AIE} develop MSE-optimal bandwidth selection and RBC inference in clustered sampling settings. \citet{Calonico-Cattaneo-Farrell-Titiunik_2019_RESTAT} investigate inclusion of pre-intervention covariates for efficiency purposes in RD designs, and develop MSE-optimal bandwidth selection and RBC inference with possibly clustered data and/or different bandwidth choices for control and treatment groups. More recently, \citet{Arai-Otsu-Seo_2021_wp} proposed novel high-dimensional methods for pre-intervention covariate selection, which lead to increased efficiency of RBC inference.

Continuity-based estimation and RBC inference methods have also been studied in other (sharp, fuzzy and kink) RD settings. For example, \citet{Xu_2017_JoE} considers generalized linear models and develops RBC inference methods, \citet{Keele-Titiunik_2015_PA}, \citet{Cattaneo-Keele-Titiunik-VazquezBare_2016_JOP,Cattaneo-Keele-Titiunik-VazquezBare_2021_JASA}, \citet{Bertanha_2020_JoE} address estimation and RBC inference in Multi-Score, Geographic and Multi-Cutoff RD designs, \citet{Dong-Lee-Gou_2021_JASA} offer a complete analysis of estimation and RBC inference in an RD design with continuous treatment, and \citet{Chiang-Hsu-Sasaki_2019_JoE}, \citet{Qu-Yoon_2019_JBES}, \citet{Chen-Chiang-Sasaki_2020_ET}, and \citet{Huang-Zhan_2021_JBES} investigate quantile RD designs and also propose RBC inference methods. 

Finally, within the continuity-based framework, scholars have addressed different failures of standard assumptions in canonical and other RD designs. \citet{Dong_2019_JBES} explores issues of sample selection in sharp and fuzzy RD designs. For fuzzy RD designs, \citet{Feir-Lemieux-Marmer_2016_JBES} and \citet{He_2018_Dissertation} study methods robust to weak identification, while \citet{Arai-Hsu-Kitagawa-Mourifie-Wan_2021_QE} consider specification testing. In addition, \citet{Pei-Shen_2017_AIE}, \citet{Lee_2017_JEM}, \citet{Davezies-LeBarbanchon_2017_JoE} and \citet{Bartalotti-Brummet-Dieterle_2021_JBES} investigate issues of measurement error in various RD settings.

%%%%%%%%%%%%%%%%%%%%%%%%%%%%%%%%%%%%%%%%%%%%%%%%%%%%%%%
%% SUBSECTION: Local Randomization Methods
%%%%%%%%%%%%%%%%%%%%%%%%%%%%%%%%%%%%%%%%%%%%%%%%%%%%%%%
\subsection{Local Randomization Methods}\label{subsec: Local Randomization Methods}

Once a local randomization assumption has been invoked, the analysis of the RD design can proceed by deploying various methods from the analysis of experiments. This requires two steps. First, a window $\W$ where the local randomization is assumed to hold must be chosen. Second, given $\W$, enough assumptions about the assignment mechanism must be imposed to perform estimation and inference, as appropriate for the specific approach chosen (Fisherian, Neyman, super-population). These methods can be used with both continuous and discrete scores (Section \ref{subsec: Discrete Score}).
 
The window selection step is the main implementation challenge for local randomization analysis, and is analogous to the bandwidth selection step in the continuity framework.  \citet{Cattaneo-Frandsen-Titiunik_2015_JCI} recommend a data-driven procedure to select the window based on pre-treatment covariates or placebo outcomes known to be unaffected by the treatment. The procedure is based on the common approach of testing for covariate balance to asses the validity of randomized experiments. If the RD local randomization assumption holds, pre-treatment covariates should be unrelated to the score inside $\W$. Assuming that the covariates are related to the score outside of $\W$ leads to the following procedure for selecting $\W$: consider a sequence of nested windows around the cutoff, and for each candidate window perform ``balance'' tests of the null hypothesis of no treatment effect, beginning with the largest window and sequentially shrinking it until the null hypothesis fails to be rejected at some pre-set significance level. The procedure thus searches subsequently smaller windows until it finds the largest possible window such that the null hypothesis cannot be rejected for that window and cannot be rejected for any smaller window contained in it. 
 
Once the window has been chosen, the second step is to deploy methods from the analysis of experiments to estimate RD treatment effects and perform statistical inference inside the selected window. These methods require to specify a randomization mechanism inside the window. A common strategy is to assume that every unit with score within $\W$ was assigned to treatment according to a fixed-margins randomization, where the probability of each treatment assignment vector is $\binom{N_{\W}}{N^+_{\W}}$, where $N^+_\mathcal{W}$ denotes the number of treatment units within the window $\W$, and $N^-_\mathcal{W}=N_\mathcal{W}-N^+_\mathcal{W}$. Another common strategy is to assume that units were assigned independently to treated and control inside the window, with probability equal to $N^+_{\W}/N_{\W}$.

Point estimation of average effects can be based on simple difference-in-means for observations inside $\W$. The sharp and fuzzy parameters, $\tau_{\mathtt{SLR}}$ and $\tau_{\mathtt{FLR}}$, can be respectively estimated with
\begin{align*}
\widehat{\tau}_\mathtt{SLR} =  \bar{Y}^+_\mathcal{W} - \bar{Y}^-_\mathcal{W} \qquad  \text{and} \qquad
\widehat{\tau}_\mathtt{FLR} = \frac{\bar{Y}^+_\mathcal{W} - \bar{Y}^-_\mathcal{W}}{\bar{D}^+_\mathcal{W} - \bar{D}^-_\mathcal{W}},
\end{align*}
where 
\begin{align*}
    \bar{Y}^+_\mathcal{W} &= \frac{1}{N_\mathcal{W}} \sum_{i:X_i\in\mathcal{W}} \frac{\omega_i}{\P[T_i=1]} T_i Y_i,\qquad
    \bar{Y}^-_\mathcal{W} = \frac{1}{N_\mathcal{W}} \sum_{i:X_i\in\mathcal{W}} \frac{\omega_i}{1-\P[T_i=1]} (1-T_i) Y_i,\\
    \bar{D}^+_\mathcal{W} &= \frac{1}{N_\mathcal{W}} \sum_{i:X_i\in\mathcal{W}} \frac{\omega_i}{\P[T_i=1]} T_i D_i,\qquad
    \bar{D}^-_\mathcal{W} = \frac{1}{N_\mathcal{W}} \sum_{i:X_i\in\mathcal{W}} \frac{\omega_i}{1-\P[T_i=1]} (1-T_i) D_i,
\end{align*}
and the $\omega_i$'s denote appropriate weights that are chosen according to the specific assumptions and framework employed. For example, if the assignment probabilities are not constant, then the observations must be weighted appropriately. In the Neyman framework with complete randomization (i.e., fixed-margins) for treatment assignment, $\omega_i=1$ and $\P[T_i=1]=N^+_\mathcal{W}/N_\mathcal{W}$ for all $i$, while under Bernoulli (i.e., coin-flip) treatment assignment, $\omega_i=1$ and $\P[T_i=1]=p$ (known) for all $i$. Finally, in the super-population framework, the previous estimators are unbiased, but in practice it is common to use the (consistent but biased) estimator that sets $\omega_i=\frac{\P[T_i=1]N_\mathcal{W}}{N^+_\mathcal{W}}T_i+\frac{(1-\P[T_i=1])N_\mathcal{W}}{N^-_\mathcal{W}}(1-T_i)$. \citet{Cattaneo-Titiunik-VazquezBare_2017_JPAM} discuss methods incorporating covariate adjustments.

Methods for inference also depend on the framework (Fisherian, Neyman, super-population) chosen for the analysis. These frameworks, which were developed for the analysis of experiments, differ in the assumptions they make about sampling and in the null hypotheses that they consider. In the Fisherian framework, the observations in the study are seen as the population of interest, not as a random sample from a larger population. As a consequence, the potential outcomes are seen as fixed, non-stochastic quantities, and the only randomness in the model stems from the random assignment of the treatment. The null hypothesis of interest is the so-called Fisherian sharp null hypothesis, under which every unit's potential outcome under treatment is equal to that unit's potential outcome under control and the treatment effect is thus zero for all units. Under this null hypothesis, the randomization distribution of the treatment assignment can be used to impute all potential outcomes for every unit and thus derive the null distribution of any test statistic of interest, which is then used to calculate exact p-values and can be extended to calculate confidence intervals under additional assumptions. Inferences are therefore finite-sample exact and do not rely on asymptotic approximations.

In the Neyman framework, the potential outcomes are also seen as non-stochastic but, in contrast to Fisher, the null hypothesis is that the average treatment effect is zero, not that the individual treatment effect is zero for every unit. Unlike Fisher's sharp null hypothesis, the Neyman null hypothesis does not allow for full imputation of all potential outcomes, so exact finite-sample inferences are unavailable. Instead, assuming that the fixed potential outcomes are drawn from a large population, law of large numbers and the central limit theorems are used to justify consistency of the usual difference-in-means estimators and validity (albeit conservative) of confidence intervals based on the Normal distributional approximation.

Finally, in the so-called super-population framework, the observations in the study are seen as a random sample taken from a larger population. The potential outcomes are therefore independent and identically distributed random variables, not fixed quantities. As in the Neyman approach, the canonical null hypothesis of interest is that the average effect is zero, and inferences are also based on the Normal distribution via large-sample approximations. Although the Neyman and super-population approaches differ conceptually, both approaches are usually implemented similarly in practice. In contrast, the implementation of Fisherian inference relies on considering all possible realizations of the treatment assignment, and is usualy implemented via permutation methods. 

Fisherian methods are particularly useful when the number observations with score within $\W$ is small, because they provide finite-sample valid inference for the sharp null hypothesis of no treatment effect under reasonable assumptions. This inference method is particularly well-suited for analyzing RD designs because it allows researchers to include only the few observations that are closest to the cutoff. As a consequence, Fisherian inference offers a good complement and robustness check to local polynomial methods (Section \ref{subsec: Local Polynomial Methods}), which typically employ a larger number of observations further away from the cutoff for estimation and inference. On the other hand, when the number of observations inside $\W$ is large enough, researchers can also use methods based on large-sample approximations within the Neyman or super-population framework. See \citet[Section 2]{Cattaneo-Idrobo-Titiunik_2022_CUP} for a practical discussion of how to use these methods for RD analysis.
 
The local randomization approach has been applied in other RD settings. \citet*{Keele-Titiunik-Zubizarreta_2015_JRSSA} consider Multi-Score and Geographic RD designs, \citet*{Li-Mattei-Mealli_2015_AoAS} analyze principal stratification in a fuzzy RD setup, \citet*{Ganong-Jager_2018_JASA} study kink RD designs, and \citet{Cattaneo-Keele-Titiunik-VazquezBare_2016_JOP,Cattaneo-Keele-Titiunik-VazquezBare_2021_JASA} investigate Multi-Cutoff RD designs. In addition, \cite{Owen-Varian_2020_EJS} study a ``tie-breaker'' design where the treatment is randomly assigned in a small neighborhood of the cutoff but follows the RD assignment otherwise, while \citet{Eckles-Ignatiadis-Wager-Wu_2021_wp} consider a setup where exogenous measurement error is assumed to induce local randomization near the cutoff. See also \citet{Hyytinen-etal_2018_QE} and \citet{DeMagalhaes-etal_2020_wp} for related work.

%%%%%%%%%%%%%%%%%%%%%%%%%%%%%%%%%%%%%%%%%%%%%%%%%%%%%%%
%% SUBSECTION: Refinements, etc.
%%%%%%%%%%%%%%%%%%%%%%%%%%%%%%%%%%%%%%%%%%%%%%%%%%%%%%%
\subsection{Refinements, Extensions and Other Methods}\label{subsec: Refinements, Extensions and Other Methods}

We review alternative approaches that have been proposed to refine, extend or generalize the standard RD estimation and inference methods discussed in the last two sections.

\subsubsection{Empirical Likelihood Methods}\label{subsec: Empirical Likelihood Methods}

\citet{Otsu-Xu-Matsushita_2015_JoE} propose an inference approach for sharp and fuzzy RD designs combining local polynomial and empirical likelihood methods, which leads to asymptotically valid confidence intervals for average RD treatment effects under an undersmoothed bandwidth choice---but a more powerful robust bias-corrected version of their method can be constructed by increasing the order of the polynomial used (Section \ref{subsec: Local Polynomial Methods}). Empirical likelihood inference is based on a data-driven likelihood ratio function, permits incorporating known constraints on parameters, and does not require explicit estimation of asymptotic variance formulas; see \citet{Owen_2001_Book} for an introduction to empirical likelihood methodology. More recently, \citet{Ma-Yu_2020_wp} investigate the higher-order properties of local polynomial empirical likelihood inference in RD designs, building on ideas in \citet{Calonico-Cattaneo-Farrell_2018_JASA,Calonico-Cattaneo-Farrell_2022_Bernoulli} for RBC inference using local polynomial methods. In addition, \citet{Otsu-Xu-Matsushita_2013_JBES} and \citet{Ma-Jales-Yu_2020_JBES} explore empirical likelihood methods for inference on a discontinuity in a density function, which has direct application both to the RD-like designs discussed in Section \ref{sec: Designs and Parameters of Interest} and to the validation and falsification methodology discussed in Section \ref{sec: Validation and Falsification}. These empirical likelihood methods are developed within the continuity framework.

\subsubsection{Bayesian Methods}

There are also several Bayesian methods for RD designs, which are developed almost exclusively within the local randomization framework. \citet{Geneletti-etal_2015_SIM} discuss a Bayesian methodology for RD designs in the biomedical sciences, focusing on how to incorporate external information in estimation and inference for RD treatment effects. Similarly, \citet{Karabatsos-Walker_2015_BookCh} and \citet{Chib-Jacobi_2016_JAE} propose alternative Bayesian models for sharp and fuzzy RD designs, paying particular attention to issues of misspecifcation of unknown regression functions. \citet{Branson-Rischard-Bornn-Miratrix_2019_JSPI} employ Gaussian process regression methods that can be interpreted as a Bayesian analog of the local linear RD estimator, thereby offering a flexible fit for regression functions of treatment and control units via a general prior on the mean response functions of the underlying Gaussian processes. Their approach is motivated by Bayesian methods, but they also discuss some frequentist properties of the resulting estimation procedures. These Bayesian methods for RD analysis rely on specifying prior distributions for functional parameters (e.g., $\E[Y_i(1)|X_i=x]$ and $\E[Y_i(0)|X_i=x]$). See \citet{Gelman-etal_2013_Book} for a general introduction to Bayesian methodology.

\subsubsection{Uniform-in-Bias Methods}\label{subsec: Uniform-in-Bias Methods}

\citet{Imbens-Wager_2019_RESTAT} and \citet{Armstrong-Kolesar_2020_QE} apply ideas from \citet{Backus_1989_GJI} and \citet{Donoho_1994_AoS} to develop RD inference methods that are uniformly valid over a smoothness class of functions determining the largest possible misspecification bias of the RD treatment effect estimator. Their procedures require pre-specifying a theoretical smoothness constant that determines the largest possible ``curvature'' of the unknown conditional expectation functions under treatment and control, $\E[Y_i(1)|X_i=x]$ and $\E[Y_i(0)|X_i=x]$, near the cutoff. Given that choice of smoothness constant, the approach computes an optimal uniform-in-bias bandwidth choice, which is then used to construct local polynomial confidence intervals for RD treatment effects that account for the worst possible bias of the RD point estimator (by inflating the quantile used). These methods are developed within the continuity framework for RD designs.

Uniform-in-bias methods depend on the theoretical smoothness constant, which is a tuning parameter that must be chosen. If this constant is estimated from data, the resulting inference procedures cease to be uniformly valid. As a consequence, assuming the goal is to achieve uniformity, the choice of the theoretical smoothness constant cannot be data-driven and must be set manually by the researcher. However, specifying this constant manually is equivalent to manually choosing the local polynomial bandwidth, because there is a one-to-one correspondence between the two parameters. Although there are some proposals for choosing the theoretical smoothness constant using data, their applicability is limited because they are either not equivariant to transformations of the data, do not employ a consistent estimator of a well-defined quantity, or are based on visual inspection of the data by the researcher.

\subsubsection{Shape-Constrained Methods}

Monotone regression estimation and inference is a useful methodology whenever a regression function is known to be monotonic because it does not require a choice of bandwidth or kernel function at interior points \citep[see][and references therein]{Groeneboom-Jongbloed_2014_Book}. Motivated by this observation, \citet{Babii-Kumar_2021_JoE} study RD treatment effect estimation and inference under the assumption that the conditional expectation functions of potential outcomes are monotonic near the cutoff. Due to the inherent boundary bias in RD settings, however, some of the nice features of monotone regression estimators are lost: a choice of tuning parameter is still required for implementation, since otherwise the monotone regression procedures would not be valid at the cutoff. Nevertheless, within the continuity-based RD framework, their shape-constrained RD methodology can be of interest in settings where monotonicity of the regression functions $\E[Y_i(1)|X_i=x]$ and $\E[Y_i(0)|X_i=x]$ near the cutoff is a reasonable assumption.

\citet{Hsu-Shen_2021_JAE} also consider the role of monotonicity in RD designs. Within the continuity-based RD framework, they propose a nonparametric monotonicity test for RD treatment effects at the cutoff conditional on observable characteristics. Their testing procedure employs local polynomial methods and therefore requires a choice of bandwidth, and their implementation relies on undersmoothing techniques, which may affect size control of the proposed test. A robust bias-corrected version of their testing procedure can be constructed by increasing the order of the polynomial used for inference (Section \ref{subsec: Local Polynomial Methods}).

\subsubsection{Other Methods}

\citet{Frolich-Huber_2019_JBES} and \citet{Peng-Ning_2021_PMLR} consider nonparametric covariate adjustments in RD designs where pre-treatment covariates are discontinuous at the cutoff, which leads to different RD-like parameters in general, while \citet{Gerard-Rokkanen-Rothe_2020_QE} propose partial identification methods in settings where the RD design is invalidated by endogenous one-sided sorting of units near the cutoff. Finally, \citet{Mukherjee-Banerjee-Ritov_2021_wp} augment the usual RD setup by including a second equation that assumes the score is a linear function of observed covariates to construct alternative semiparametric estimators of average RD treatment effects.

%%%%%%%%%%%%%%%%%%%%%%%%%%%%%%%%%%%%%%%%%%%%%%%%%%%%%%%
%% SUBSECTION: Departures
%%%%%%%%%%%%%%%%%%%%%%%%%%%%%%%%%%%%%%%%%%%%%%%%%%%%%%%
\subsection{Discrete Score Variable}\label{subsec: Discrete Score}

The defining feature of the RD design is the treatment assignment rule $\I(X_i \geq c)$, which in principle allows for both a discrete or continuous score $X_i$. A continuously distributed score implies that all units have a different score value in finite samples, and ensures that some observations will have scores arbitrarily close to the cutoff in large samples (provided the density of the score is positive around the cutoff). In contrast, a score with discrete support implies that multiple units will share the same value of $X_i$, leading to repeated values or ``mass points'' in the data. In the absence of a continuously distributed score, standard continuity-based RD treatment effects are not nonparametrically identifiable at the cutoff---this problem is equivalent to the lack of nonparametric identifiability in bunching designs \citep{Blomquist-Newey-Kumar-Liang_2021_JPE}. The main issue is that, with discrete scores, identification and estimation of continuity-based RD treatment effects would necessarily require extrapolation outside the support of the score.

There are different approaches for RD designs with discrete scores, depending on the assumptions imposed and the conceptual framework under consideration. To describe these methods, suppose $X_i$ is discrete and can take $M = 2K+1$ distinct values $\{x_{-K}, \ldots, x_{-2}, x_{-1}, c, x_1, x_2, \ldots x_K \}$, where we assume the cutoff $c$ is the median value only for simplicity. Within the continuity-based framework, the local polynomial methods discussed in Section \ref{subsec: Local Polynomial Methods} can still be employed when (i) the implicit parametric extrapolation from $X_i=x_{-1}$ to $X_i=c$ is accurate enough, and (ii) the number of unique values, $M$, is large. In such settings, conventional bandwidth selection, estimation, and inference methods might still be applicable. On the other hand, when the score has only a few distinct values, continuity-based methods are not advisable for analysis of RD designs unless the underlying local polynomial regression model for the conditional expectations is believed to be correctly specified. See \citet*{Lee-Card_2008_JoE} for early methods under this type of assumption.

Local randomization methods, on the other hand, remain valid under the same assumptions discussed in Section \ref{subsec: Local Randomization Methods} when the score is discrete, and the estimation and inference methods in that section can be applied regardless of whether the number of distinct values of $X_i$ is large or small. If the number of observations per mass point is large enough, the window selection step is not needed, as it suffices to use observations with $X_i = x_{-1}$ and $X_i = c$ to conduct estimation and inference using Fisherian, Neyman, or super-population methods. If the number of observations per mass point is not large enough so that a bigger window is needed, the window selection can be based on pre-determined covariates, but the implementation is easier because the nested windows considered can be increased in length one mass point at the time on either side of the cutoff.

In some cases, it may be desirable to redefine the parameter of interest. For example, within the continuity-based framework, the average RD treatment effect $\tau_\mathtt{SRD}=\E[Y_i(1)-Y_i(0)|X_i = c]$ is not identifiable without extrapolation (from the left of the cutoff); but the parameter $\tau_\texttt{SDS}=\E[Y_i(1)|X_i = c] - \E[Y_i(0)|X_i = x_{-1}]$ is always identifiable, making explicit the inability to identify and estimate $\E[Y_i(0)|X_i = c]$ without additional modelling assumptions. The new parameter $\tau_\texttt{SDS}$ may be more natural in cases where the score is inherently discrete, but its usefulness is ultimately application-specific.

In sum, from a practical perspective, a key consideration for RD analysis with discrete scores is the number of distinct values $M$ in the support of the running variable. When $M$ is large, continuity-based methods may still be useful, though the effective number of observations for the purposes of local polynomial estimation and inference will be the number of mass points (not the total number of units). When $M$ is small, continuity-based methods are only useful under additional assumptions for extrapolation, and local randomization methods may be a preferable alternative. See \citet[Section 4]{Cattaneo-Idrobo-Titiunik_2022_CUP} for more details and a practical discussion.

There are other specific approaches for RD settings with discrete scores. \citet{Dong_2015_JAE} and \citet*{Barreca-Lindo-Waddell_2016_EconInq} investigate the phenomenon of heaping, which occurs when the score variable is rounded so that units that initially had different score values appear in the dataset as having the same value. \citet*{Kolesar-Rothe_2018_AER} employ a uniform-in-bias method, which requires specifying a theoretical bound on the ``curvature'' of a class of regression functions that are only identifiable at the mass points of the score; see also Section \ref{subsec: Uniform-in-Bias Methods}.

\subsection{Power Calculations for Experimental Design}

The discussion so far has focused on ex-post analysis of RD designs using either continuity-based or local randomization methods. In recent years, applied researchers and policy-makers have been increasingly interested in designing ex-ante program evaluations leveraging RD designs, where researchers have access to the scores of all units but must choose a subset of units for whom to collect outcomes due to budget considerations or other restrictions. This experimental RD setting requires computing power functions for RD hypothesis tests, which depend on the specific estimation and inference methods adopted. \citet{Schochet_2009_JEBS} discusses power calculations and optimal sample size selection for experimental design using parametric regression specifications for RD treatment effect estimation. Within the continuity-based framework, \citet{Cattaneo-Titiunik-VazquezBare_2019_Stata} develop power calculations and compute optimal sample sizes for ex-ante experimental design when employing RBC inference (Section \ref{subsec: Local Polynomial Methods}). These power calculation results can also be used to compute Minimum Detectable Effects (MDE), which are useful for both ex-ante and ex-post analysis of RD designs. \citet{Bulus_2021_JREE} derives MDE formulas for clustered RD designs under parametric assumptions. Importantly, for ex-post analysis, computing MDE effects is preferred because ex-post power calculations using observed effect sizes are often unreliable.

%%%%%%%%%%%%%%%%%%%%%%%%%%%%%%%%%%%%%%%%%%%%%%%%%%%%%%%
%% SECTION: Validation and Falsification
%%%%%%%%%%%%%%%%%%%%%%%%%%%%%%%%%%%%%%%%%%%%%%%%%%%%%%%
\section{Validation and Falsification}\label{sec: Validation and Falsification}

Relative to other non-experimental designs, the RD design offers a wealth of design-specific validation and falsification methods to assess the plausibility of the core RD assumptions. Regardless of whether the continuity or local randomization framework is used, a fundamental step in any RD analysis is to provide such supporting empirical evidence. This evidence can only be indirect, because the core RD assumptions are inherently untestable. Nonetheless, there are important steps that can be taken to assess their plausibility in most settings, and hence the potential validity of the RD design, as well as the overall robustness of the main empirical findings. We first discuss generic falsification methods that are used in most program evaluation settings, and then turn to design-based methods that rely on specific RD features. See \citet*[Section 5]{Cattaneo-Idrobo-Titiunik_2020_CUP} for a recent practical overview and more details.

\subsection{Pre-intervention Covariates and Placebo Outcomes}

In randomized experiments, it is common to check whether the randomization produced similar distributions of observed covariates for control and treatment groups, where the covariates are determined before the treatment is assigned. This so-called \textit{covariate balance} analysis is usually implemented by testing the null hypothesis that the mean (or other features of the  distribution) of pre-determined covariates is the same in the group of units assigned to treatment and the group of units assigned to control. Rejecting the null hypothesis implies that the treatment and control groups are not comparable in terms of pre-treatment characteristics, thereby casting doubts on the research design. The same idea can be applied in the RD design \citep{Lee_2008_JoE}. In addition, the principle of covariate balance can be extended beyond pre-determined covariates to variables that are determined after the treatment is assigned but are known to be unaffected by the treatment, commonly referred to as ``unaffected'' \citep[][p. 121]{Rosenbaum_2010_Book} or ``placebo'' outcomes. This kind of validation analysis can be powerful, as it relies on the hypothesized mechanism by which the treatment affects the outcome. For example, in their analysis of Head Start expansion on child mortality, \citet[][Table III, p. 180]{Ludwig-Miller_2007_QJE} estimated the RD effect on causes of death that should not have been affected by Head Start programs and found such effect to be indistinguishable from zero, in addition to also considering other pre-intervention covariates.

Analysis of pre-intervention covariates and placebo outcomes is a fundamental component of the validation and falsification toolkit for RD designs. Its application is straightforward: local polynomial RBC inference methods (Section \ref{subsec: Local Polynomial Methods}) and local randomization inference methods (Section \ref{subsec: Local Randomization Methods}) can be used directly upon substituting the outcome variable of interest by the pre-intervention covariates or placebo outcome. In the usual implementation of balance tests, failure to reject the null hypothesis of covariate balance is interpreted as evidence of comparability of control and treatment groups near the cutoff. However, it may be more appropriate to test the null hypothesis that the covariate distributions are different and thus control the probability of concluding that the groups are similar when they are not (rather than the probability of concluding that they are different when they are similar, as in the usual approach). See \citet{Hartman_2020_PA} for an application in RD settings using RBC inference in the continuity-based framework.

Finally, other methods for pre-intervention covariates and placebo outcome testing have been proposed within the continuity framework. \citet{Shen-Zhang_2016_RESTAT} propose local polynomial distributional tests using an undersmoothed bandwidth for implementation, but a more robust version using RBC inference can be constructed by increasing the polynomial order (Section \ref{subsec: Local Polynomial Methods}). \citet{Canay-Kamat_2018_RESTUD} discuss an approach based on asymptotic permutation tests, which leads to a randomization inference implementation (Section \ref{subsec: Local Randomization Methods}).

\subsection{Density Continuity and Binomial Testing}

\citet{McCrary_2008_JoE} proposed a density test for score ``manipulation'' near the cutoff, where the null hypothesis is that the density function of the score is continuous at the cutoff. While continuity of the density function of the score at the cutoff is neither necessary nor sufficient for validity of an RD design, a discontinuous density can be heuristically associated with the idea of ``endogenous sorting'' of units around the cutoff. Furthermore, under additional assumptions, such discontinuity can be formally linked to failure of the RD design. For example, \citet{Urquiola-Verhoogen_2009_QJE} and \citet{Bajari-Hong-Park-Town_2011_wp} study economic models that predict endogenous sorting around the cutoff, while \citet{Crespo_2020_JCI} discusses sorting caused by administrative reasons rather than strategic individual behavior.

McCrary's density test has become one of the leading validation and falsification methods for RD design with a continuously distributed score. The test was originally implemented using the local polynomial density estimator of \citet{Cheng-Fan-Marron_1997_AoS}, a two-step nonparametric procedure requiring two distinct bandwidth choices. More recently, \citet{Cattaneo-Jansson-Ma_2020_JASA} introduced a local polynomial density estimator that requires only one bandwidth choice and is more robust near boundary points, which leads to a hypothesis test based on RBC inference, thereby offering an improved density test for RD applications.

Building on McCrary's idea, but originally motivated by the local randomization framework, \citet{Cattaneo-Titiunik-VazquezBare_2017_JPAM} propose a binomial test for counts near the cutoff as an additional manipulation test. Unlike the continuity-based density test, the binomial test can be used when the score is continuous or discrete, and does not rely on asymptotic approximations, as the associated p-value is exact for a given assignment probability in a fixed window around the cutoff. This validation method offers a useful complement and robustness check for the continuity-based density test.

Within the continuity-based framework, there are other extensions and refinements for density testing. \citet*{Otsu-Xu-Matsushita_2013_JBES}, \citet*{Ma-Jales-Yu_2020_JBES} and \citet*{Jales-Ma-Yu_2017_EL} discuss density testing combining local polynomials and empirical likelihood methods; see Section \ref{subsec: Empirical Likelihood Methods}. \citet{Frandsen_2017_AIE} proposes a manipulation test for discrete scores and \citet{Bugni-Canay_2021_JoE} propose a binomial-like test using an asymptotic argument to justify local randomization near the cutoff, both employing a uniform-in-bias approach that requires the manual choice of a theoretical smoothness constant; see Section \ref{subsec: Uniform-in-Bias Methods}.

\subsection{Placebo Cutoffs and Continuity of Regression Functions}

Another falsification method specific to the RD design looks at artificial (or placebo) cutoffs away from the real cutoff determining treatment assignment. The intuition is that, since the true cutoff is the only score value at which the probability of receiving the treatment changes discontinuously, it should also be the only score value at which the outcome changes discontinuously. This test assumes that there are no other policies (or changes) introduced at different score values that may affect the outcome of interest. More formally, this method can be viewed as hypothesis test for continuity of the conditional expectation functions of potential outcomes at values of the score equal to the placebo cutoff considered. For implementation, researchers usually choose a grid of artificial cutoff values, and repeat estimation and inference of the RD effect on the outcome of interest at each artificial cutoff value, using either local polynomial RBC inference methods (Section \ref{subsec: Local Polynomial Methods}) or local randomization inference methods (Section \ref{subsec: Local Randomization Methods}). Importantly, in order to avoid treatment effect contamination, this validation method should be implemented for units below and above the cutoff separately. See \citet{Ganong-Jager_2018_JASA} for a related approach in the context of kink RD designs.

\subsection{Donut Hole and Bandwidth Sensitivity}

Another class of RD-specific falsification methods involves re-implementing estimation and inference for the RD treatment effect with different subsets of observations, as determined by either excluding those observations closest to the cutoff or varying the bandwidth used for estimation and inference. The so-called ``donut hole'' approach \citep*{Barreca-Guldi-Lindo-Waddell_2011_QJE} investigates the sensitivity of the empirical conclusions by removing the few observations closest to the cutoff, and then implementing estimation and inference. Heuristically, the goal is to assess how much influence the few observations closest to the cutoff have on the overall empirical results. The intuition is that if there is endogenous sorting of units across the cutoff, such sorting might occur only among units whose scores are very close to the cutoff, and thus when those observations are excluded the RD treatment effect may change. More formally, this approach can be related to issues of local extrapolation properties of the specific method used for estimation and inference near the cutoff. See \citet{Dowd_2021_wp} for the latter approach in the context of local polynomial RBC inference methods (Section \ref{subsec: Local Polynomial Methods}).

The other validation method studies whether the empirical conclusions are sensitive to the choice of bandwidth or local randomization neighborhood. The idea is simply to re-estimate the RD treatment effect for bandwidths (or neighborhood lengths) that are smaller or larger than the one originally chosen. In the continuity-based framework, if the original bandwidth is MSE-optimal, considering much larger bandwidths is not advisable due to the implied misspecification bias. Similarly, in the local randomization framework, considering larger neighborhoods may not be justifiable if important covariates become imbalanced; in this case, the approach will be uninformative.

For implementation of the donut hole and bandwidth sensitivity approaches, researchers can use either local polynomial RBC inference methods (Section \ref{subsec: Local Polynomial Methods}) or local randomization inference methods (Section \ref{subsec: Local Randomization Methods}).

%%%%%%%%%%%%%%%%%%%%%%%%%%%%%%%%%%%%%%%%%%%%%%%%%%%%%%%
%% SECTION: Conclusion
%%%%%%%%%%%%%%%%%%%%%%%%%%%%%%%%%%%%%%%%%%%%%%%%%%%%%%%
\section{Conclusion}\label{sec: Conclusion}

Since the seminal work of \cite{Thistlethwaite-Campbell_1960_JEP}, the literature on identification, estimation, inference, and validation for RD designs has blossomed and matured. Thanks to the collective work of many scholars, current RD methodology provides a wide range of both practical and theoretical tools. The original analogy between RD designs and randomized experiments has been clarified and in many ways strengthened, the choice of local polynomial bandwidth and local randomization neighborhood has become data-driven and based on principled criteria, and many validation and falsification methods have been developed exploiting the particular features of RD designs. It is now understood that best practices for estimation and inference in RD designs should be based on tuning parameter choices that are objective, principled, and data-driven, thereby removing researchers' ability to engage in arbitrary specification searching, and thus leading to credible and replicable empirical conclusions.

Looking ahead, there are still several areas that can benefit from further methodological development. One prominent area is RD extrapolation. Having more robust methods to extrapolate RD treatment effects would bolster the external validity of the conclusions drawn from RD designs, which are otherwise limited by the local nature of the parameters. Such extrapolation methods would also help in the design of optimal policies, a topic that has not yet been explored in the RD literature. Another fruitful area is experimental design and data collection in RD settings. It is becoming more common for policy makers and industry researchers to plan RD designs ex-ante as well as to design ex-post data collection exploiting prior RD interventions, often in rich data environments. A set of methodological tools targeted to experimental design in RD settings would improve practice and could also enable better extrapolation of RD treatment effects via principled sampling design and data collection. Finally, another avenue for future research is related to incorporating modern high-dimensional and machine learning methods in RD settings. Such methodological enhancements can aid in principled discovery of heterogeneous treatment effects as well as in developing more robust estimation and inference methods in multi-dimensional RD designs.

%%%%%%%%%%%%%%%%%%%%%%%%%%%%%%%%%%%
%%%%%%%%%%%%%%%%%%%%%%%%%%%%%%%%%%%
%%%%%%%%%%%%%%%%%%%%%%%%%%%%%%%%%%%
\newpage
\section*{References} %Remove the asterisk to get References in the TOC, with a numbered section.
%\small
\singlespacing
\begingroup
\renewcommand{\section}[2]{}	%this removes the standard 'References' header that BibTeX puts in.
\bibliography{CT_2022_ARE--bib}
\bibliographystyle{econometrica}
\endgroup

\end{document}